\documentstyle[aps,epsfig]{revtex}
\tightenlines
\begin{document}
\author{G.J.A. Sevink$^*$}
\address{Leiden Institute of Chemistry,
Leiden University,
PO Box 9502, 2300 RA Leiden, The Netherlands}
\author{A.V. Zvelindovsky}
\address{Centre for Materials Science, 
University of Central Lancashire, Preston, PR1 2HE, United Kingdom}
\title{Block copolymers confined in a nanopore: Pathfinding in a curving and frustrating flatland.}
\maketitle

\begin{abstract}
We have studied structure formation in a confined block copolymer melt by means of  
dynamic density functional theory (DDFT). The confinement is two-dimensional,
and the confined geometry is that of a cylindrical nanopore.
Although the results of this study are general, our coarse-grained molecular model is 
inspired by an experimental lamellae-forming PS-PBD diblock copolymer system 
(Shin {\it et al}, Science, {\bf 306}, 76 (2004)), in which an exotic toroidal structure 
was observed upon confinement in alumina nanopores. 
Our computational study shows that a zoo of exotic structures can be formed, 
although the majority, including 
the catenoid, helix and double helix that were also found
in Monte Carlo (MC) nanopore studies, 
are metastable states. We introduce a general classification scheme
and consider the 
role of kinetics and elongational pressure on 
stability and formation pathway of both equilibrium and metastable structures in detail.
We find that helicity and three-fold connections 
mediate structural transitions on a larger scale. Moreover, by matching the remaining parameter
in our mesoscopic method, the Flory-Huggins parameter $\chi$, to the experimental
system, we obtain a structure that resembles the experimental toroidal structure in great detail.
Here, the most important factor seems to be the roughness of the pore, i.e. small
variations of the pore radius on a scale that is larger than the characteristic size in the system.
 
\end{abstract}

\section{Introduction}
Pattern formation of
block copolymers in constraint situations or {\it confinement} is an important topic in polymer
research, since the meso- or microstructure can be much better controlled when compared to
the bulk. In general, molecular conformations and assembly are strongly influenced 
by confinement. In the absence of external constraints, the microstructure
is dictated by the interaction between
segments comprising the copolymer, the volume fraction
of the blocks, and the molecular architecture. 
In a confined system, however, interfacial interactions, symmetry
breaking, structural frustration and confinement-induced entropy loss
play a determining role and may lead to structures 
that differ from the ones found in bulk. 
As a direct result phase separation in confinement has been the subject of
extensive theoretical and experimental studies.\cite{wangbook} 
The intriguing prospects
from a technological viewpoint are the novel structures that can
be achieved and that may serve as scaffolds for other nanostructures.\\
The simplest example of a block copolymer is a linear AB 
diblock copolymer.  In the bulk several stable periodic microstructures can be formed, 
among which are lamellar, hexagonal and body-centered cubic phases.\cite{fre99} 
The equilibrium behavior for AB diblock copolymers in the bulk
has been mapped out both experimentally and
theoretically; the situation for more complex multi-block
and/or branched block copolymers is much less clear.
Here, we focus on symmetric or nearly symmetric AB diblock copolymers. 
In the bulk, these copolymers microphase separate
into lamellar microdomains with a characteristic equilibrium period $L_0$, where
grains of ordered lamellar microdomains are randomly oriented. 
Global orientation of the microdomains can be induced by confinement of the block copolymer
in thin supported films or slits (one dimensional confinement).
\cite{matsen98,binder99,fasolka01,hashimotobook} Additionally, the interplay
between surface fields (interaction with confining surfaces) 
and confinement effects (commensurability) can affect the phase behavior,
and lead to the formation of surface reconstructions or hybrid structures.
\cite{radzilowski96,huinink00,wang01}\\  
A obvious next step in experimental and theoretical research
is to consider systems where the confinement is effectively two-dimensional. This is the case when 
the melt is confined inside a cylindrical nanopore of radius $R$. From a conceptual
point of view, this type of confinement differs from the 1D confinement
since: (1) there is only one confining surface, so there is only one
surface field, (2) the nanopore solid surface is curved. By definition,  
surface reconstructions (analogs of the uni- or multilayer parallel lamellae $L_{||}$ in slits) 
are therefore curved as well, which has an effect on 
the entropy contribution of the structure in the free energy. 
This can either result in a break-up of the nanopore in two relatively independent regions (close
and away from the pore surface), or in curved multilayer structures that experience 
lamellar bending throughout the pore, with an increased bending towards
the centre of the pore, (3) just like in slits, surface reconstructions may be able to adapt their
layer spacing to some extent in order to suit the cylinder 
radius $R$. One can expect the influence of this type of frustration to
be less than in slits, as there is no block-preference in the centre of the pore.
However, for small $R$, structures can experience substantial frustration due
to incommensurability. To resolve this unfavorable situation, the lamellae can adapt an orientation
perpendicular to the nanopore wall or the chains may find alternative packing and 
form hybrid structures, like in a slit, and (4)
in addition to the commensurability issue for parallel structures or surface reconstructions (see (2)) 
the length of a nanopore can also affect the formation of perpendicular 
structures (incommensurability along the pore). 
In slits, these perpendicular structures are found for strong surface fields at 
incommensurable film thickness, and for weak surface fields independent of the film thickness. 
In nanopores, a mismatch of the pore length and the natural spacing $L_0$ 
can lead to extensional forces, and may stabilise hybrid structures as well.\\ 
Based on rather small data set of surface interactions
and nanopore radii $R$, two pioneering
computational studies (by Monte Carlo\cite{he01} and dynamic self-consistent field
theory\cite{Sevpore}) identified two structures:
a slab and a multiwall tube morphology.\cite{Sevpore} 
In analogy with the classification scheme for structures in a slit,\cite{our_prl,fasolka00}
slabs can be associated with lamellae perpendicular to
the confining surface (L$_{\perp}$) and multiwall tubes
with surface reconstructions: lamellae that are parallel to the confining surface (L$_{||}$).
One should note that this remark implies that the nature of these structures
is the same as the lamellar structure found in bulk.  It was concluded 
that two mechanisms control the structure formation in a nanopore :
(1) in the case of weak surface interactions 
the lamellae orient perpendicular 
to the pore wall to form lamellar slabs 
and (2) for strong surface interactions, one of the blocks
segregates preferentially to the pore wall, and the lamellae line
up and bend to form concentric cylinders, the number of which
is determined by the cylinder radius. The  
effect of commensurability was found to be less significant in this type of confinement, since 
no perpendicular structures were observed
for incommensurate $R$ and higher surface interactions.\\
Very recently, the significance of these theoretical studies
increased dramatically by the appearance of a number 
of intricate experimental studies of polystyrene-{\it block}-polybutadiene (PS-{\it b}-PB) 
\cite{shin04,xiang04,wu04,xiang05}
and polystyrene-{\it block}-poly(methyl methacrylate) (PS-{\it b}-PMMA)\cite{sun05} diblock
copolymers confined in nanopores. In these studies,
bulk lamella-, cylinder- and sphere-forming block copolymers were introduced 
into nanoscopic cylindrical pores in alumina membranes.
For bulk lamella-forming block copolymers the predicted morphology \cite{he01,Sevpore}
of slabs was not found, most probably due to the strong disbalance of 
surface-energies in the experimental
setup (although these were not measured as such). The most frequent found structure
was a concentric cylinders (multiwall tube \cite{Sevpore}) morphology. 
However, also a new structure, a stacked-disc or toroidal-type \cite{shin04}
structure, was found inside a nanopore with an incommensurate pore diameter 
($d/L_0 \sim 2.6$, with $d$ the pore diameter).\\
As a result of these experimental findings, computational
studies have considered the phase behaviour of confined symmetric 
diblock copolymers in more detail, by means of self-consistent field theory (SCFT) \cite{li06} 
and Monte Carlo (MC)\cite{chen06,feng06a,feng06b,wang07} methods. Based on 2D 
calculations, the work of Li {\it et al} \cite{li06} aimed at constructing
phase diagrams ($\chi N$ versus $f=f_A$), and found lamellae for $f_A=1/2$
(the diagram was for calculated for a single radius $R=8.5 R_g$). Although
this work can be seen as a stepping stone for the understanding of the
nature of phase transitions under varying conditions, their findings do not directly
relate to the essentially 3D structures found in experiments. For instance, the
slab morphology is out of the scope of their calculations. Moreover, the influence
of the nanopore radius and the surface energetics was not considered in detail. Based on lattice
MC simulations, Chen {\it et al}, Feng {\it et al} and Wang indeed found 
several new structures in 3D: a single helix, catenoid cylinder, 
gyroidal, stacked circle, and disordered structure in Ref. \cite{chen06},
a mesh, lamellae parallel to the pore axis, 
single helix and double helix structure in Ref. \cite{feng06a} 
and catenoid cylinder in Ref. \cite{wang07}. A later
study of Feng {\it et al} concentrated on the stability of the helical structure.\cite{feng06b} 
From a visual comparison we conclude that the mesh structure\cite{feng06a} 
is the same as catenoid cylinder,\cite{chen06}. 
Similarly, one can claim that the gyroidal structure is in fact a defected structure,
and comprises a coexistence of a mesh and a double helix structure.\cite{chen06}
We adapt a common notation, and conclude that this reduces the number of 
computed nanopore structures to seven: stacked disc, concentric cylinder, lamellae parallel
to the pore axis, catenoid cylinder, disordered, single and double helix.\\
Although new structure were found, a fundamental understanding of the
underlying mechanisms such as achieved for thin films
(effective 1D confinement) is absent, except for the MC study of Wang \cite{wang07}. 
However, the latter study is restricted to strong segregation, and most remarkably several of the other 
structures (for instance the interesting helix and double helix) were not found at all.
Moreover, the origin of the experimental 
stacked-disc or toroidal-type \cite{shin04} structure is still unexplained. 
More fundamental studies, including scans of a larger parameter space, are 
clearly needed for a deeper understanding.\\
In principle, one can anticipate several regimes. 
Equilibrium morphologies are minima of the free energy, containing 
both energetic and entropic contributions.
In the absence of any surface field and for a strong surface field, 
either one of these contributions is dominating, leading
to the stacked disc and concentric cylindrical structures. 
Between these regions, for weak surface fields, the chains have more flexibility to adapt
their packing and the system can and apparently will adapt other morphologies. 
Incommensurability may play a subtle role here. 
The details of this interplay remains to be determined,
for instance by the calculation of a structure diagram in 3D, 
depending on $R$ and the surface interaction strength. Since we consider
lamella-forming systems, $f_A$, the volume fraction of the $A$ block, 
is fixed (in contrary to Ref \cite{li06}) and
the diagram is conceptually similar to the one on Ref.\cite{our_prl} for confined films.\\  
Moreover, two factors have not been considered in detail yet: the
influence of the kinetic pathway and the value of $\chi_{AB} N$, where $N$ is
the total length of the block copolymer.
Although we will not focus into detail in the latter factor, from Ref \cite{li06}
one can conclude that an increase of $\chi_{AB} N$ may lead to a decrease of the
number of concentric lamellae.
Here, we focus on calculating the structural diagram and the question of stability.
Whether interesting structures like double helix can actually be manufactured 
experimentally may be a very subtle issue. It is known that in strong
confinement the dynamics may slow down and 
structures may be frozen into metastable states.
This is particularly the case when the free energy difference
between states is small or when there is a large
energy barrier between different local minima of the free energy.
In nanopores, this situation was actually 
observed in MC studies: in Ref. \cite{chen06} 
for weak surface fields up to three different structures were found.
Previously, a dynamic density functional theory (DDFT) 
study \cite{Sevpore} identified the single helix structure
as a long living intermediate structure between complete mixing and
stacked discs for zero surface field. 
We will use this method here to consider 
the experimental system of Ref. \cite{shin04}
in detail.
In contrast to traditional schemes of polymer phase separation
dynamics where a Landau Hamiltonian is used with vertex functions
calculated following the Random Phase Approximation (see {\it e.g.}
\cite{THE}), we numerically calculate the free energy $F$ of
polymer system consisting of Gaussian chains in a mean field environment
using a path integral formalism \cite{fraaije97,vlimmer99}. Our approach
uses essentially the same free energy functional as in SCF
calculations of equilibrium block copolymer morphologies by Matsen and
Schick \cite{matS}, but complements the static SCF calculations by
providing a dynamical picture of the system.

\section{Method}
Here, we shortly discuss the DDFT method \cite{fraaije97,sevink99} 
for a bulk lamella-forming diblock copolymer melt. The 
diblock copolymers are modelled by a $A_{N_A} B_{N_B}$ Gaussian chain
($N=N_A+N_B$, $f_I=N_I/N$).
The confined geometry is a cylindrical pore with varying diameter $R$. Periodic
boundary conditions apply in the $x$-direction, along the pore. Calculations are
carried out on a cubic $L_x \times L_y \times L_z$ grid with a spacing $\Delta x$ that is related to the
Gaussian bond length $a$ via $a \Delta x^{-1}=1.1543$.\cite{maurits96a}
The spacing $\Delta x$ is equal among different pore systems, making the free energy per
volume element of $\Delta x^3$ easily comparable. Unless mentioned otherwise,
all spacings are in units of this basic variable $\Delta x$. The pore is 
introduced into the simulation volume $V$ by a masking technique; 
as a result the simulation volume contains both pore and mask points.\cite{sevink99}
In mask points, elements of the subset
$V^0=\{{\bf r}=(x,y,z)\in V| \|(x,y,z) - (x,y_0,z_0)\|>R\}$ with $y_0=L_y/2$, $z_0=L_z/2$, 
all concentration and external potential 
field values are set to zero, except for the
auxilary field $\rho_M({\bf r})=1$ that represents the mask itself.\cite{sevink99}
The free energy for unconfined systems 
is given by (see \cite{fraaije97} for details) 
\begin{eqnarray}
F[\rho]&=&-kT\ln \frac{\Phi^n}{n!}-\sum_I\int_V U_I({\bf r})\rho _I({\bf r})d{\bf r}\nonumber\\&+&\frac{1}{2}\sum_{I,J}\int_{V^2}^{}\varepsilon_{IJ}(|{\bf r}-{\bf r'}|)\rho_I({\bf r})\rho_J({\bf r'})d{\bf r}d{\bf r'}\nonumber\\&+&\frac{\kappa}2\int_V\left( \sum_I \nu (\rho _I({\bf r})-\rho_I^0)\right)^2d{\bf r}\;,
\label{free_en}
\end{eqnarray}
Here, $k$ is the Boltzmann constant, $T$ is the 
temperature, $n$ is the number of polymer molecules in the volume $V$ 
occupied by the system,  
and $\Phi$ is the intra-molecular partition function for ideal polymer chains. 
The parameter $\kappa$ determines the compressibility of the system 
(the dimensionless $\kappa'=\beta \kappa \nu=20.65$, with $\nu$ the bead volume), 
and $\rho_I^0$ is the mean concentration of the $I$-block (where the average is taken over 
$V \setminus V^0$). For $\kappa \rightarrow \infty$ the system becomes 
incompressible. The external potentials $U_I$ and the concentration fields $\rho_I$ 
are related via the density functional.\cite{vlimmer99} 
The inter-chain interactions are incorporated
via a mean-field with interaction strength controlled by the Flory-Huggins parameters $\chi_{IJ}$. 
In line with our earlier work the interactions are
specified by the parameters 
$\varepsilon^0_{IJ}$ (in kJ/mol), \cite{fraaije97,vlimmer99}
which are directly related to the Flory-Huggins parameters by $\chi_{IJ}=
1000 \varepsilon_{IJ}^0/n_A k T$ (with $n_A$ Avogadro's number
and $T=300$ the temperature in Kelvin). 
In case of non-zero surface interactions an extra
cohesive term is added to the free energy (\ref{free_en}) equal to \cite{sevink99}
\begin{equation}
F^{wall}=\sum\limits_{I}^{} \int\limits_{V^2} \varepsilon_{IM}(|{\bf r}-{\bf r}'|) \rho_I({\bf r})
\rho_M({\bf r'}) d{\bf r} d{\bf r}'\;.
\label{wallterm}
\end{equation} 
In line with our earlier work, the interaction kernel is chosen Gaussian, and
the important parameter $\epsilon^0_{IM}$ 
denotes the scalar interaction strength \cite{fraaije97} of bead $I$ with the pore boundary.
In the computations, $\varepsilon^0_{BM}=0$ indicating that the $B$-blocks have no 
interaction with the wall, which is appropriate due to the fact
that for an incompressible system of diblocks only the effective surface interaction 
$\xi=\varepsilon^0_{AM}-\varepsilon^0_{BM}$ is of importance. Using this we obtain
\begin{equation}
F^{wall} = \sum\limits_{I}^{} \varepsilon^0_{IM} {\cal F}[\rho_I,\rho_M]= \xi {\cal F}[\rho_A,\rho_M]
\label{wallen}
\end{equation}
where ${\cal F}$ is 
\begin{equation}
{\cal F} =  {\cal C} \int\limits_{V^2}^{} 
e^{-\frac{3}{2a^2}(r-r')^2} \rho_A({\bf r}) \rho_M({\bf r}') d{\bf r}d{\bf r}'
\end{equation}
with ${\cal C}=(\frac{3}{2 \pi a^2})^{\frac{3}{2}}$ a normalisation constant and $a$ the Gaussian
bond length. Since the bead-bead interaction $\varepsilon^0_{AB}$ is mostly
considered a constant, we will use the notation $(R,\xi)$ to denote
points in the structural diagram.\\
For the simplest model, the evolution of the density fields  
is given by a Langevin equation \cite{maurits97}
\begin{equation}
\frac{d{\rho _{I}}({\bf r})}{dt}=M\Delta \mu_I({\bf r}) +\eta _{I}({\bf r})
\label{kinconst}
\end{equation}
with $M$ a constant mobility, $\Delta$ the Laplace operator and
$\eta$ noise, distributed according to the fluctuation-dissipation theorem.
Other transport coefficients with a more
physical scaling behaviour, like the one for 
collective Rouse dynamics or reptation,
exist, \cite{maurits97} but are in general too computationally demanding. 
We have recently found that our relatively simple 
model (\ref{kinconst}) 
can be appropriate to describe the experimental dynamics in detail.
In Ref. \cite{ting} the experimental and calculated dynamics of
cylinder and sphere forming diblock copolymers under an external electric field was compared, and
good agreement was found based on DDFT with constant transport coefficients. 
In Ref. \cite{our_natmat} the calculated dynamics of DDFT was shown to  
agree in full detail with SFM measurements of experimental dynamics 
in a thin film of a concentrated SBS solution. 
Finally, apart from the extra term in the free energy, confinement is accounted for by 
the boundary condition for the
dynamic equations $n \cdot \nabla \mu_I=0$, with $n$ the normal
pointing into the solid object. \cite{sevink99} 

\section{Results and Discussion}
\subsection{Nanopores: data from literature.}
Since the nanopore simulation data in literature is rather scattered,
\cite{he01,Sevpore,chen06,feng06a} we first present a short overview.
A direct comparison of these different studies is
complicated by the fact that important system parameters are different. 
A Monte Carlo (MC) method was used to simulate the 
phase behaviour for a $A_5B_5$ diblock copolymer \cite{feng06a} with
$\varepsilon_{AB}=0.3, 1.0$ and $1.1$ kT, and for a $A_{10}B_{10}$
diblock copolymer \cite{he01,chen06} with $\varepsilon_{AB}=1.0$ kT. In both
studies all other interactions are zero: only the $\varepsilon_{AS}$ (in kT), the
interaction between A-blocks and the surface, is
varied. In the dynamic density functional theory (DDFT) calculations 
of Ref.\cite{Sevpore}, a $A_8B_8$ system
was considered for $\varepsilon_{AB}=2.5$ kJ/mol and varying $\varepsilon_{AS}$ (in kJ/mol)
(in that article, $S$ is actually called $M$ or mask). Using the formula in the method section 
for $T=300 K$, we obtain $\chi=1.0$ and $\chi N = 16$ (weak segregation). 
Using the expression from Ref.\cite{freire03}, $\chi \approx 5 (\varepsilon/kT)$,
we can recalculate the parameters used in the MC studies:
$\chi N \approx 15, 50$ and $55$ (weak to intermediate segregation)
for Ref. \cite{feng06a} and $\chi N \approx 100$
in Ref. \cite{he01,chen06} (strong segregation), where it should 
be noted that Ref. \cite{feng06a} concentrates
on $\chi N \approx 15$ (weak segregation). As a consequence, the results
are distributed between weak, intermediate and strong segregation 
regimes. In Figure \ref{fig1} we have combined the existing knowledge in
schematic diagrams for different $\chi N$; we have only differentiated between 
stacked discs, concentric cylinders and alternative structures. It should be noted that 
the diagram for DDFT is essentially no {\it{phase diagrams}}, as
the final structures are pathway dependent, and not minima of the free energy per se.
This is a fundamental difference
between static calculations aimed at deriving equilibrium morphologies,
and our dynamic simulations aimed at mimicking experimental pathways, including
visits to long-living metastable states.
The important dimensionless 
spatial coordinate is the ratio of the pore diameter $R$ and the 
lamellar domain distance $L_0$; this
parameter is on the vertical axis. 
On the horizontal axis is $\varepsilon_{AS}$ (in kT) of MC. Since the diblock is
symmetric we can restrict ourselves to only positive values.
We have used $\varepsilon_{AS}$ (in kT) $\approx 1/2 \varepsilon^0_{AS}$
(in kJ/mol) for the conversion of the DDFT to the MC value.\\
We see that concentric cylindrical structures
dominate all diagrams when $\varepsilon_{AS}$ is large, 
irrespective of the value of $\chi N$.
In this case, the surface field dominates and one block is energetically favoured 
at the pore wall.
For neutral pores, i.e. in the absence of a surface field ($\varepsilon_{AS}=0$), stacked 
discs dominate. In stacked discs the chains
can adopt a packing that is similar to the one in normal bulk lamellae. 
However, for the strongest segregated system stacked disc coexist with a 
single helix for this surface field. For small non-zero values
of $\varepsilon_{AS}$ also other structures can be found.\\
Although the data are sparse, one can 
conclude that structures with alternative packing dominate 
the diagram for large $\chi N$
and relative small $\varepsilon_{AS}$. Incommensurability cannot be the only 
important factor since these alternative structures
are even found for commensurate pore radii $R/L_0=1$, \cite{chen06} although
there is some uncertainty since the bulk domain distance $L_0$ was not determined explicitly in this
study. We can only conclude from these data that especially weak surface fields in combination with 
strong segregation (high $\chi N$) lead to structures with alternative packing.

\subsection{Nanopores: system choice and boundary condition}
\label{bulk}
The system was chosen to model the experimental polystyrene-{\it block}-polybutadiene (PS-{\it b}-PBD)
diblock copolymer in Ref. \cite{shin04}, which has a volume fraction of 0.56 for the
butadiene block. The molecular model considered here is a $S_{10} B_{12}$
($N_S=10$, $N_B=12$) Gaussian chain. The interaction between $S$ and $B$ beads
is chosen as $\varepsilon_{SB}=2.0$ kJ/mol or $\chi=0.8$, and consequently $\chi N = 17.6$. 
A diblock copolymer melt of this molecular composition $f_B=0.545$
forms lamellae in bulk, and the bulk lamellar distance was determined as $L_0=8.6$ 
(from here onwards, all distances are in units of $\Delta x$). 
Commensurability issues in the $x$ direction, along the pore axis, may arise as a result of
periodic boundary conditions. Since the effects of periodic boundary conditions
have been considered in the literature,\cite{wang00,morita04} we refer to these works
for a detailed discussion. Here, we only note that the studies mentioned in the
previous paragraph have not considered this effect in detail.
Only for one dataset in Ref. \cite{chen06}, were both helices and 
stacked discs were found and no transition between the two different structures,
it was shown that the number of MC simulations required to first find a helical structure peaks at 
particular cylinder lengths (our $L_x$). General conclusions from this study
are not easy to make due to the absence of regularity.

\subsection{Nanopores: results and discussion}
We have considered pattern formation in a 
$S_{10} B_{12}$ diblock copolymer melt confined in a nanopore of
length $L_x=32=3.72 L_0$, except for the larger pores where we
have used $L_x=34=3.95 L_0$ and $L_x=36=4.19 L_0$.
In order to give a unifying description for the structural behaviour 
of slightly asymmetric diblock copolymers in nanopores, we first
construct and discuss a diagram of simulated structures. Since these structures
are the result of dynamic pathways and therefore
may be metastable, we consider
their stability by an interpolation procedure described below. Consequently, 
we focus on several factors that are important for structure formation:
two types of {\it incommensurability} issues, due 
to packing frustration along the pore (associated with perpendicular structures)
and perpendicular to the pore  
(associated with  parallel structures), and 
the kinetic pathway, that may lead to arrested structures. 
In the second part, we discuss these issues in detail, and unravel the
complex interplay of these factors and the surface field.\\
Packing frustrations along the pore, originating from the requirement of periodicity
at the two pore boundaries, have not received much attention
in the past, except for the
unconfined situation.\cite{morita04} In this study, it was shown that this frustration, quantized by $L_x/L_0$, 
determines the (in)stability of several perpendicular structure. This effect, that can be related
to {\it elongational stress or extensional force} in an applied shear field,
will be discussed in detail for our results.\\
In the absence of extensional forces, the analogy to similar substrates in thin films suggests 
that the phase behavior in pores
is due to an interplay of two factors: the strength of the surface field and confinement 
effects.\cite{wangbook,our_prl,huinink01}
For vanishing or weak surface fields,
the elastic chain deformation associated with parallel structures will be avoided,
and perpendicular structures are favored instead.
Due to the selective block-surface interaction,
parallel structures (surface reconstructions) will be promoted 
for surface fields above a certain threshold field strength.
However, the available space in
the confined geometry dictates the degree of chain compression 
or extension required for parallel structures. 
Large chain frustrations give rise to an entropically 
unfavorable situation and may even prevent the formation of parallel structures.
Alternative (in thin films, perpendicular) structures are then formed instead.
In thin films, the surface fields were shown to be additive and affect only the first layers of structure.\cite{huinink01} 
Consequently the available space, quantized by $R/L_0$ in pores,
is important for both factors.
The surface field and frustration due to confinement will strongly influence structure formation in small pores.
In larger pores, the surface field has a limited range, 
and possible structure frustration can be distributed over more layers of structure.\\
We note that the diblock copolymer considered here is slightly asymmetric. 
As a consequence, the structure diagram is not completely symmetric as well. 
In the absence of surface interactions,
the shortest ($S$) part of the chain is preferentially 
found close to the surface due to entropic effects. The value for which
energetic and entropic contributions are balanced 
shifts to a small positive $\xi$.\cite{our_prl}
Moreover, for parallel structures in the curved geometry, 
this chain asymmetry leads to {\it curvature effects}, since the 
spontaneous curvature associated with each of the different blocks is slighlty different,
and {\it asymmetric packing frustrations} due to the different $S$ and $B$ domain sizes. 
In contrast to thin films, the majority component of the centre layer in a parallel structure is not prescribed 
by surface energetics. Packing frustrations are therefore expected to be 
less significant than in thin films, where in most cases the block next to the surfaces
is prescribed by the surface fields.
However, the number of domains/layers for each of the blocks can {\it differ}.
To focus on this effect we adapt two notations: 
if the majority component in a parallel structure is $S-B-S-B-S$ or $B-S-B-S-B$,
on a line perpendicular to the pore axis from wall to wall, we denote the structure as 
$L_{||,1 \frac{1}{2}}$ (for a concentric cylindrical structure, in line with the slit notation \cite{our_prl}),
and by the order of the majority component in the layers
from wall to centre, $SBS$ or $BSB$. Note  
that here the total number of $S$ and $B$ domains differ (2 or 3), 
as well as the composition of the central cylinder, associated with the largest curvature.
Curvature is an important factor and a complicating factor, when compared to the thin film situation.
In principle, each of the layers in a parallel structure can 
adapt their thickness to some extent in response to global mismatches.   
Simple volumetric arguments indicate that the domain spacing in nanopore confinement 
depends on the local curvature, and therefore 
on the absolute radial position of $S-B$ interfaces in the pore.\cite{wang07}
This is particular the case in the limit of strong segregation; for weaker segregation,
the situation may be different, since the blocks are somewhat miscible.
As a result, the spacing may heterogeneously
deviate from the bulk domain spacing ($L_0=8.6$, so $D_S=3.9$ and $D_B=4.7$), 
near the pore centre, where the curvature is highest, and in the wetting layer,
where confinement effects are most severe.
For a symmetric diblock copolymer and larger radii $R$, we previously showed that 
packing frustrations in parallel structures are mainly relieved by rearrangement
in the cylindrical centre region; in the layered structure away from the centre, the chain conformations 
are rather unaffected by the local curvature. \cite{Sevpore}
We will consider this interplay of surface field and frustration in detail for our asymmetric block copolymer.\\ 
We have varied the surface interaction in a range of
negative and positive values $\xi$, 
between vanishing (non-selective) 
to intermediate (selective, $S$ or $B$) surface fields. 
A range of small $R$ up to approximately $L_0$ was chosen to consider the details of this interplay 
in strong confinement. Moreover, also a few larger $R$ were considered to study
these effects separately. The structures are displayed 
in Figures \ref{fig5}-\ref{fig7}, and the free energies associated with these
structures in Figure \ref{fig8}. In the selected region in the $(R,\xi)$ diagram, we expect that
confinement effects, surface fields and/or the interplay between these two
gives rise to alternative structures. 
In the remainder, we use the relevant dimensionless parameter $R/L_0$ 
instead of the bare radius $R$.
For all calculated final structures, the free
energy was monitored and remained constant.\\
We introduce a short-hand notation for the perfect structures: 
catenoid cylinder are denoted by $PL$ or $PL(I)$ (depending on the sign of $\xi$, the 
majority component $I$ of the structure is either $S$ (negative) or $B$ (positive)), disordered by $D$, 
single helix by $H$ and double helix by $DH$. The lamellae parallel to the cylinder axis are
perpendicular to the pore wall but also parallel, and we denote them as $L_{\perp,||}$.
In some cases, defected structures are remarkably stable. Based on visual inspection,
and knowledge about metastable intermediate states along the pathway of formation
of the perfect structures, we assign a symmetry group 
preceded by the letter '$d$' (for instance, $dDH$ is a defected double helix).
Other new structures will be discussed and annotated as they appear.\\
Focussing on the kinetic pathway, we remark that all structures in the diagram of Figure \ref{fig5} 
were obtained following a diffusive pathway
(equation (\ref{kinconst})). Earlier work showed that microphase separation 
in confined systems often starts close to the pore 
wall.\cite{lyakhova06} For the pore dimensions considered here,
this effect will be small, but, especially for stronger surface fields,
surface reconstructions will initially form, resulting in overall coverage of the pore wall
by a single component. The transition to more stable structures with a different 
surface coverage (for instance, helices or
stacked circles) requires considerable transport of material away for the wall, 
into the center of the nanopore. The fundamental mechanism of this transition is
important, as it may hint when and where this process may be kinetically trapped.
First, we consider which of the calculated structures is an equilibrium structure.
For varying $\xi$, the free energy (see Figure \ref{fig8}) associated with a particular structure 
(defined by $\rho_S$; due to incompressibility, $\rho_B$ is then also fixed)
is given by (see also (\ref{free_en})) 
$F^{tot}=F+F^{wall}=F[\rho_S]+\xi {\cal F}[\rho_S,\rho_M]$.
Since both $F[\rho_S]=a$ and ${\cal F}[\rho_S,\rho_M]=b$ have a constant value for
a particular $\rho_S$ ($\rho_M$ is the mask field, and fixed by definition) the free energy values
for this structure and varying surface field can be found on a line defined by $a+\xi b$. 
We only need two datapoints (or the values of $a$ and $b$) for the same structure 
to determine this line. However, we have to be careful using this procedure
as structures with the same morphology {\it type} may differ in detail (and therefore in the 
values of $a$ and $b$), such as geometrical quantities and the degree of 
segregation.
For instances, interpolation suggests that the $L_{\perp}$ 
structures for $\xi=-0.4,-0.3$ and $\xi=0.1,0.2$ (both for $R=0.70 L_0$) are different, 
although there is no structural difference in terms of easy computable geometrical quantities.
Only when we consider the bare density values we find that the 
maximum of the concentration field $\theta_S({\bf r})$ for $\xi=-0.4,-0.3$ is slightly 
higher than for the equivalent structures at $\xi=0.1,0.2$.

\subsubsection{Structure diagram and stability}
The structure diagram is shown in Figure \ref{fig5}, and is indeed not symmetric. 
Equilibrium structures are distinguished by a grey background;
this determination is based on the interpolation approach described above.
Considering the general features of this diagram,
one observes that several structures in this range are metastable.
For large $R/L_0$, only $L_{||}$ and $L_{\perp}$ are stable; for small $R/L_0$ also
other structures can be stable. 
The effect of incommensurability is limited, since we do not observe outliers of 
perpendicular structures for the larger radii 
considered (where the effect of the surface field is relatively small). 
However, for positive $\xi$ (pore surface likes $B$) and $R/L_0=1.28$ or $2.21$, 
the $L_{||}$ structure 
is only metastable and the transition to equilibrium structures trapped. 
Overall, this suggests
that the pore radii $R/L_0=1.28$ and $2.21$ are incommensurate, 
while $R/L_0=1.74$ is commensurate.
Moreover, we see that for negative $\xi$ (pore surface likes $S$), 
$L_{||}$ is stable for $\xi \leq -0.4$, independent of the radius $R/L_0$. 
This asymmetry is due to the curvature effects and asymmetric packing 
frustration mentioned above.\\
Focussing on specific structures, we see that for large absolute values of $\xi$ surface 
reconstructions, $PL$, $dPL$, $dL_{||}$, $L_{\perp,||}$ and $L_{||}$, dominate. 
In particular, 
we find $PL$ and $L_{||,1}$ ($SB$ or $BS$) for $R/L_0 \in \{0.58,0.70\}$,
$L_{||,1 \frac{1}{2}}$ ($SBS$ or $BSB$) for $R/L_0 \in \{0.81,0.93,1.05,1.28\}$,
$L_{||,2}$ ($SBSB$ or $BSBS$) for $R/L_0=1.74$, and
$L_{||,2 \frac{1}{2}}$ ($SBSBS$ or $BSBSB$) for $R/L_0=2.21$.
An analysis of domain spacing is given in Table I; the values
are derived from the interface locations.
Disconnected concentric cylinders $dL_{||}$, where the cylinder close to the pore surface
is broken up into 4 disconnected parallel lamellar patches, 
only appear for $R/L_0=0.93$. The distance
between the patches increases with increasing $\xi$ in this region. 
The $L_{\perp,||}$ is the same as the structure in Fig 8b in Feng {et al},\cite{feng06a} 
and related to $dL_{||}$ (see discussion lateron).
The number of perforations in $PL$ can vary and show hexagonal ordering when 
considered in the 2D plane. Geometrically, the $PL$ structure can be seen as 
an intermediate between
$L_{||}$ and $H$.  
Defected perforated lamellae $dPL$, distinguished from $PL$ since 
the perforations do not show ordering on a larger scale, are found on the
boundary between parallel and perpendicular structures.  In response to the perforations
the central cylinder sometimes adapts an oval cross-section. Defected $PL$
only appear for negative $\xi$ (surface preference to $S$). 
The majority of surface reconstructions ($PL$ or $L_{||}$)
for positive $\xi$ (surface preference to $B$) are not stable structures.
They are examples of kinetically trapped structures, due to 
the presence of the pore surface, that gives 
rise to $L_{||}$ related structures in the early stages of phase separation.\\
Perpendicular structures, titled tacked discs
$L^{tilt}_{\perp}$ and $L_{\perp}$ (4 discs for all $L_x$ considered), are found in the centre of the diagram, for small $\xi$.
For small relative radii $R/L_0$, the discs are perpendicular to the pore surface, $L_{\perp}$.
For larger $R/L_0$ there are only three discs, and they are tilted with 
respect to this surface, $L^{tilt}_{\perp}$. For the largest pore radii, $R/L_0=1.74$ and $2.21$, 
the pore length $L_x$ ($L_x=34 = 3.95 L_0$ and $L_x=36 = 4.19 L_0$, respectively)
is somewhat larger than the standard $L_x=32$, and the structure is $L_{\perp}$. 
The $L^{tilt}_{\perp}/L^{tilt}_{\perp}$ structure are two coexisting
perpendicular $L^{tilt}_{\perp}$ structures, with opposite tilt angles. The details of
these mixed structures, not representing equilibrium structures for obvious reasons, will
be considered later.\\
Helical structures, $H$, $DH$ and a new $GH$ structure, 
are found at the rather broad boundary between perpendicular and parallel structures.
These structures possess
features relating them to both perpendicular and parallel structures: the axis of winding is parallel
to the pore axis, and the wall-coverage is nonuniform. 
The $GH$ structure in $(1.05,0.4)$ is related to the helical structures, but differs topologically since 
small helical patches are three-fold connected with helical patches on the opposite side, and 
these connections join into 
two cylinders parallel to the pore axis. The $GH$ for $(1.16,0.5)$ and $(1.40,0.4)$
share this property, but instead the connections form two lamellar 
patches and three cylinders, respectively.
We call this structure gyroid-helical ($GH$) because of the three-fold connectivity.
Stable helices (left- and righthand)
and a double helix are formed for $R/L_0 < 1$ and negative $\xi$. The
helical structures for positive $\xi$ are all metastable. Metastable $GH$ are
only found for larger effective radii, $R/L_0>1$, adjacent to both parallel and perpendicular
structures.\\
Several coexisting structures, $L_{\perp}/PL$, $L_{\perp}^{tilt}/H$, $L_{\perp}^{tilt}/GH$, 
$L_{\perp}^{tilt}/H$ and $L_{\perp}/L_{||}$, and
defected structures, $dL_{\perp}^{tilt}$, $dL_{\perp}$, $dGH$, $dL_{\perp,||}$
are found directly adjacent to their 'perfect' counterparts, and stay
defected after many TMS. Apparently, the driving force for the
removal of different types of defects in these structures is rather 
small, and as a result the structures are kinetically trapped. The only
exceptions, coexisting $H/PL$ and $PL/L_{||}$ for positive $\xi$ that are
not adjacent to a $PL$ structure, show that $PL$ is associated
with a free energy close to the one for $H$ and $L_{||}$, respectively. For $R/L_0=0.81$
and $-0.4 \leq \xi \leq 0.0$ the structures are very defected, and remain as such,
even after a large number of extra timesteps (TMS).  

\subsubsection{The value of $L_x/L_0$: elongational stress and perpendicular structures}
Earlier work for a 'soft' confined system\cite{morita04} concluded that the 
$L_{\perp}$, $L_{\perp}^{tilt}$ and $H$ structures are related. Their stability
depends on the extensional force on the system, originating either from
an external field (experiments) or from boundary conditions (computations). 
Extensional forces may be present in experimental and computational studies
dealing with structure formation in nanopores,
except for a computational study in the strong segregation limit (SSL),\cite{wang07} 
where MC based on a grand canonical ensemble was employed, but have
not received much attention.
Especially in computational methods considering a canonical ensemble and a single $L_x$, 
this type of commensurability issues cannot always be avoided.
We argue that a deeper understanding of this effect is relevant. In experiments, 
for instance, extensional forces play a role when the pore surface is 
very rough, the pore length is very small (often the case in applications considered in soft 
nanotechnology), and when shear fields are present, as is the case in 
a recently developed experimental technique for the fabrication of nanowires.\cite{ma06}
In the present study, we have considered this effect by varying $L_x$ for a small set of selected parameters:
$L_x=34=3.95 L_0$ for $(0.58,-0.1)$ and $(1.28,0.1)$ and $L_x=36=4.19L_0$
for $(0.70,-0.5)$, $(0.81,0.2)$, $(0.93,0.2)$ and $(1.05,0.1)$.
In all cases the microstructure evolves into $L_{\perp}$ (4 discs), with an associated free energy
that is lower than for the orginal structure for $L_x=32$.
A detailed analysis of the structure evolution revealed
that $L_{\perp}^{tilt}$ and $H$ structures can also be
kinetically related to $L_{\perp}$: in some cases, they mediate large-scale structural reorganisation.  
This phenomenon can be observed from the helicity that appears during the evolution of the $L_{\perp}$ structure 
in a pore of $L_x=36$ and $(R/L_0,\xi)=(0.70,-0.5)$ (see figure \ref{fig10} for
the details of the formation pathway for different $L_x$). It is in 
agreement with earlier findings for a fully symmetric $AB$ diblock in Ref. \cite{Sevpore}.\\
Apart from the single additional calculation for each of pores and $L_x \neq 32$, 
we used the $L_{\perp}$ structures to compute the free energy 
for an additional surface field strength $\xi$. Interpolation between these two values enables us 
to reconsider the stability of the structures in Figure \ref{fig5} with respect to $L_{\perp}$. 
We make two general remarks:  
i) Although the interpolation technique is valuable for comparing the stability of different
microstructures in pores of {\it equal} length (for varying $\xi$), 
in principle this procedure cannot be used to differentiate
between microstructures in pores of {\it different} length. 
For instance, an instantaneous change of the pore length (or equal, extensional force) during the
evolution gives rise to a deformation of the structure, and possibly a transition to a more stable structure.
However, in contrast to self-consistent field (SCF) techniques, DDFT does not impose symmetry,
and changes in free energy associated with a deformation of an existing structure,
for instance an affine deformation of a helical structure,
cannot be quantized directly. 
Instead, additional calculations are necessary for many alternative structure types
and are very time consuming in general.\cite{morita04}
We therefore anticipate on the results of the SSL study, 
\cite{wang07} that considered nanopore structure formation in the absence of elongational 
stress. This study identified, 
besides lamellar structures, only stable $PL$.
Consequently, we adapt an practical approach and use 
all data obtained by simulation (see Figure \ref{fig8}), independent of $L_x$,
to suggest a phase diagram of equilibrium structures
(see Figure \ref{newfigure}), without considering
the deformability of the non-lamellar structures. 
We note that the stability of the $PL$ phase could not be determined indefinitely, 
since the alternative $L_{||}$ phase was not formed for these
small radii. Only for $R/L_0=0.58$ and positive $\xi$, concentric cylinders ($BS$) were formed, and
the region where $PL$ is equilibrium structure could be determined. 
ii) The additional simulations for $L_x \neq 32$ were carried out for 
relatively weak surface fields, where perpendicular structures are likely to form. 
Interpolation indicates (see figure \ref{newfigure}) that the equilibrium structure can be perpendicular
even for relatively strong surface fields,
for instance due to incommensurability along the radial direction.
However, the large structural rearrangements required for the transition
to (stable) perpendicular structures, starting from the (unstable) parallel structures that are 
initially formed due to the strong surface interaction, 
may lead to arrested structures, i.e. microstructures that are kinetically trapped in metastable states.
A good example is the $DH$ structure in $(0.81,-0.5)$. Although the phase diagram of Figure \ref{newfigure}
shows that $\L_{\perp}$ is the equilibrium structure, we
challenged the stability of the $DH$ structure by choosing a number of systems
with different pore length $L_x$ (each system was quenched from a homogeneous mixture at 
TMS=0, see Figure \ref{fig11}-\ref{fig12}). 
We never obtained the equilibrium structure as a result of the dynamic pathway. Instead, upon a variation of
the extensional pressure, we obtain structures with an
overall parallel orientation, that lack symmetry on a larger scale. Only for $L_x=32,35$ and §$38$
we find an almost perfect $DH$ structure. We rank these other structures in three different classes,
labelled as $DHi$. Structure $DH1$ can be seen as a highly interconnected and
defected $L_{\perp,||}$ structure, where the centre lamellae contains large holes and is connected
to the other structure. Structure $DH2$ is completely symmetric; the axis of symmetry is
$L_x/2$. Structure $DH3$ consist of two disconnected sheets of different structure,
one similar to $PL(B)$ and the other similar to $PL(S)$. For $L_x=64$ the structure is 
very defected and a combination of other structures. 
Comparing the free energies in Figure \ref{fig12}, we find that the one of the metastable
$DH$ structure ($L_x=32$) is the lowest, 
indicating that the extensional force is minimal for $L_x=32$ and
the structure is trapped due to the diffusive kinetic pathway. 
The finding of $DH$ for other $L_x$ shows that the factor $3/L$ (with $L$ the particular spacing of the structure) 
plays an important role. In general, we see that an 
increase of the nanopore length slows down the separation dynamics.
Close examination of the structures in diagram \ref{fig5} reveals that the calculated structures for $-0.4\leq \xi \leq 0.0$ are
defected $DH$ or resemble the intermediate $DH1$. 
In all cases, the kinetics disables the mass transport necessary for 
the $DH \rightarrow L_{\perp}$ transition.

\subsubsection{The value of $R/L_0$: commensurability and parallel structures}
From the phase diagram in Figure \ref{newfigure} (see discussion for the derivation of this
diagram above) we distinguish two specific values: $R=0.81 L_0$
and $R=1.74 L_0$. For $R=0.81 L_0$, the pore radius $R<L_0$ and the surface field is therefore strong, but
confinement effects apparently prevent the formation of parallel structures, in favor of
the $L_{\perp}$ structure, up to large surface field stengths. Since this radius also marks the transition of $SB$ 
($0.70 L_0$) to $SBS$ (or $BS$ to $BSB$), we conclude that $R=0.81 L_0$
is incommensurate. For $R=1.74 L_0 > L_0$, the surface field influence is much less. Apart from a symmetric
$L_{||}$ region for large surface fields (we note that the number of $S$ and $B$ domains is the same, independent
of the wetting block), only a small
region of perpendicular structures is observed for almost neutral pores, and we conclude that the pore size is commensurate.
We consider the simulation results in Figure \ref{fig5} in more detail. Domain distances are shown in Table I,
and are inexact for the $PL$ structure, which is not radial symmetric.
For the smallest $R=0.58 L_0$, parallel structures are found for strong surface fields.  
Most of these structures are $PL$ ($PL(S)$ or $PL(B)$), and $L_{||}$ is only found 
for $(0.58,0.7)$. The interpolation procedure showed that, at the onset of the $L_{||}$ region, the free energies associated
with both parallel phases
are relatively close (the value of $b$ is almost equal, see discussion of the interpolation procedure before),
which explains that $L_{||}$ is found for $\xi=0.7$ due to kinetic trapping.
Upon comparing the $PL$ and $L_{||}$ structures geometrically, we see that 
the formation of necks on the central cylinder in $PL$
leads to an increased curvature of the $S-B$ interface
and contact area of the non-preferred block with the pore surface. 
If we compare the position of this $S-B$ interface for this incommensurate and 
the commensurate situation ($R/L_0=1.74$, $S-B$ interface closest to the pore centre, see Table I) 
we observe a reduced thickness for
positive $\xi$ and equal thickness for negative $\xi$. 
This illustrates the curvature effect mentioned before, meaning that the 
elastic chain deformation associated with the formation of $L_{||}$ can 
much easier be facilitated when the interface is curved 
towards the shortest $S$-part of the chain, than to longest $B$-part.
The same phenomenon can be observed for $R=0.70 L_0$. Here,
concentric cylinders ($L_{||}$) are completely absent in the simulated range. Alternative 
$PL$ structures, very similar to the ones for $R=0.58 L_0$ apart from
the number of perforations, can be found, but equilibrium $PL(S)$ structures
are only found for rather low $\xi$ values, again due to the curvature effect. 
In order to fill the pore, the domains are rather extended (Table I).
The formation
of parallel structures requires strong stretching of the part of the chain
that contains the wetting blocks, giving rise to an entropic penalty that is
only counterbalanced for stronger surface repulsion.  
For $R=0.81 L_0$ surface reconstructions $L_{||,1 \frac{1}{2}}$ ($SBS$ or $BSB$) are found 
and $PL$ is absent, apart from the coexisting $H/PL$ structure ($\xi=0.5$). 
However, from the phase diagram, we conclude that all surface reconstructions $L_{||}$ are metastable.  
and formed due to kinetic factors. The analysis of the domain 
distances in Table I shows rather small domain sizes, and therefore strong
compression of the chains.
Although $L_{||,1\frac{1}{2}}$ is only found for positive $\xi$ ($R=0.93 L_0$), we observe many other 
parallel structures, suggesting that this radius $R$ is somewhat commensurate.
However, from the analysis in Table I we see that the domain spacing
for this radius is still rather small (compared to bulk spacing, and
the spacing of $D_3+D_4=9.0$ found for the inner two layers in the 
commensurate $BSBS$ structure 
for $\xi=0.7$ and $R/L_0=1.74$), especially towards the center. 
Morover, the equilibrium structure in this range is $L_{\perp}$ (see Figure \ref{newfigure}).
We focus on the disconnected lamellar patches ($dL_{||}$) for $\xi \leq -0.2$. Because of the
masking procedure, the cross section of the pore is not completely circular, and
the curvature of the pore wall varies. Hence, one could expect that metastable concentric
cylinders and formed due to kinetic factors, and that the origin of these disconnections,
appearing in regions of high curvature, is the chain-stretching required for the formation
of undefected $L_{||}$. Close examination of the kinetic pathway towards 
the $L_{\perp,||}$ structure (figure \ref{fig9}d) shows $dL_{||}$ as an intermediate structure. 
In particular, along the pathway two of the lamellar patches and the central cylinder have merged to form 
the centre lamellae. In the early stage, undulations lead to connections between
the different structural elements in $dL_{||}$ (very fast). Subsequently, these connections
merge (and others are removed) by a complex process that includes
the formation of holes (very slow). The structures along this pathway resemble
the metastable DH1 structures that are found in $(0.81,-0.5)$ for varying $L_x$.
Finally the $L_{\perp,||}$ structure forms. Apparently the removal of defects depends
on the surface field strength and is slowed down considerably by an increase of $\xi$ from $-0.1$ to $0.0$. 
In MC calculations\cite{feng06a} the $L_{\perp,||}$ structure was also found 
for a pore radius $R \approx L_0$. It was suggested 
that it represents an intermediate between the $L_{||}$ and $L_{\perp}$
structures, but no further explanation was given. The stability
analysis suggests that all non-perpendicular structures are metastable.  
We claim that this structure represents a kinetically arrested alternative for stable $L_{\perp}$.
Both structures are formed for small to vanishing surface fields,
in order to avoid the elastic chain deformation due to the curved $S-B$ interfaces in 
concentric cylinders.
Following the same reasoning, the $dL_{||}$ structures (instead of $L_{||}$) for $\xi<-0.1$ are
metastable structures along the pathway to this $L_{\perp,||}$ structure, and kinetically 
trapped since the undulations necessary for the formation of this structure are surpressed
by the strong surface field.\\ 
For $R=1.05 L_0$ the situation is weakly incommensurate, and $L_{\perp}$ is the
equilibrium structure between $-0.4 \leq \xi \leq 0.8$.
This can also be seen from the slightly perturbed spacing
in Table I for $\xi=-1$.  
For $-0.5 \leq \xi \leq -0.3$ in Figure \ref{fig5} ($dPL$) holes are formed in the region of highest curvature,
and we find that the density field in the centre of the 
pore is slightly higher for $\xi=-0.5$ than for $\xi=-1$ (no holes). 
This $dPL$ structure was also observed as the initial stage in the formation of
$DH$, $H$ and $L_{\perp}$ structures for these $\xi$ and smaller $R$, and for the slow
transition to $dL^{tilt}_{\perp}$ for $\xi=-0.2,-0.1$, where we simulated up to 40000 TMS. 
For positive $\xi$, the parallel structures are also metastable, but the free energy associated 
with the $PL(B)/L_{||}$ coexisting structure ($\xi=0.6$) is slightly below the one extrapolated for perfect $L_{||}$.
This is once more an indication that the chain deformation associated with concentric cylinders 
can be relieved by the formation of necks, and therefore $PL$ is more stable.
Moverover, this transition from $L_{||}$ to $PL$ (and eventually the transition, via other
intermediate structures, to $L_{\perp}$) is obviously very slow.
The parallel structure $L_{\perp,||}$ is absent, showing that the kinetic pathway leading
to this metastable structure is rather unique.\\
For $R=1.28 L_0$, the phase diagram suggests that the pore size is slightly too large for $L_{||,1\frac{1}{2}}$.
The detailed analysis in Table I shows that 
the chains in the center of the pore are stretched, compared to
the bulk domain thicknesses ($D_B=4.7$ and $D_S=3.9$), and 
to the $L_{||,1\frac{1}{2}}$ structures for other $R$.
The boundary for the equilibrium $L_{\perp}$ phase shifts
to stronger surface fields, but this shift is small compared to $R=1.05 L_0$. 
The larger shift for positive $\xi$ is due to the curvature
of the $S-B$ interface of the central cylinder, which is curved towards
the shortest block for negative $\xi$.\\
For $R=1.74 L_0$ surface reconstructions $L_{||,2}$ are found for positive and negative $\xi$. Moreover,
only at the boundary with defected structures, these surface reconstructions
are metastable. The stability region of $L_{||}$ is almost symmetric around $\xi=0.1$,
and we conclude that the situation is commensurate. 
The chains are slightly stretched (see Table I) 
compared the spacing to the domain spacing in the bulk due to curvature effects.
A comparison of the spacings in pores for $R/L_0=1.74$ and $2.21$ suggests that
especially the chains in the centre of the pore are compressed for the largest pore.
The driving force for the transition of the mestastable surface reconstructions 
to $L_{\perp}$ is rather small (see Figure \ref{fig8}). 

\subsubsection{Detailed kinetic pathways}
Although some kinetic pathways have been considered in the previous subsections,
we focus on a few more representative pathways here.
For $R=0.70 L_0$, different kinetic pathways were found leading to the final $L_{\perp}$ 
structures for $\xi=-0.4$ (figure \ref{fig9}a) and
$\xi=0.1$ (figure \ref{fig9}b). For the negative $\xi=-0.4$
the initial structure is $L_{||}$. At later stages (200 TMS) the cylinder in the centre
of the pore has broken up
to form short and disconnected cylinders, while part of the outer curved lamellae
stays intact. Remarkably, parts of this outer curved lamellae remain intact
during the transition and mediate the mass transfer
from the wall into the pores interior.
This is the source of the toroidal structures that appear as intermediate,
at the positions where the stacked discs are later formed. For the positive $\xi=0.1$
the initial structure is bicontinuous (or very defected $PL(S)$) 
with many three-fold connections in the centre
of the pore. These connections can be seen as defects, as very early in the evolution there
is an excess of material at positions where stacked discs are later formed.
Four-fold connections (small lamellar patches) and helical domains mediate
the mass transfer needed for the formation of perfect stacked discs in this case.\\
The most interesting structure for $R=0.81 L_0$ is the double helix ($DH$) for $\xi=-0.5$.
Figure \ref{fig9}c shows that the formation of this structure is a rather slow process,
and starts from $L_{||}$ (100 TMS). In the early stages, equidistant perforations 
form in the outer curved lamellae, similar to the ones formed for $R/L_0=1.05$ and 
$-0.4 \leq \xi \leq -0.3$. These perforations grow in time while the cylinder in the centre of the pore
undulates and breaks up to form a bicontinuous network. In contrast to before this bicontinuous network
itself forms the nucleus of the final structure, which is the double helix structure. Due to its
helical nature, defect removal cannot be mediated by the formation and breakage of
helical connections and thus the process is rather slow.\\
The metastable parallel structure $GH$ is only formed
for positive $\xi$, and either stabilized by an extensional force or kinetically trapped
along the pathway of the $L_{||} \rightarrow L_{\perp}$
transition. Its finding  
adjacent to perpendicular structures (deformed $L_{\perp}$) and $L_{||}$ for $R=1.28 L_0$, or perpendicular structures and
$PL$, coexisting with $L_{||}$ and $H$, for smaller pores
suggests that it may evolve from $PL$. Close examination of the pathway
for $R/L_0=1.28$ ($\xi=0.4$) and $R/L_0=0.93$ ($\xi=0.3$) shows that the initial structure
$L_{||}$ rapidly transforms (after 100 TMS) into a very defected structure, where the perforations in the outer
$B$ layer are uneven and randomly positioned.
Since $PL$ only appears in coexistence with other structures for positive $\xi$, we conclude that the 
kinetic pathways for the formation of $GH$ and $PL$ must be different.\\
For $R=1.28 L_0$, we concentrate on the effect of elongational stress on the evolution of 
the $L^{tilt}_{\perp}/L^{tilt}_{\perp}$ structure,
where the equilibrium structure is $L_{\perp}$.
In particular, we consider the structure evolution for $\xi=0.2$ in detail, since
incommensurability issues with respect to the pore radius are absent.
First, we note that the
cylinder in the centre of the pore for $\xi=-0.2$ show some undulations, but the number of pores in the outer layer 
remain constant for longer simulation times. and no connections with the inner structures
are formed within the limits of simulation time. These connections are required for the 
transition/nucleation of one phase (in)to the other. 
For $\xi=-0.1$ a defected structure without apparent symmetry at TMS=10000 evolves 
into a helical phase at TMS=25000. Since the axis of winding is away from the 
centre, we conclude that the structure can be seen as a 
majority $L_{\perp}^{tilt}$ under some angle (with the pore wall), 
connected to a minority $L_{\perp}^{tilt}$ under a reverse angle. As a natural consequence,
the connections between these two partial structures have a helical nature. We observe
similar structures for $\xi=0.1$ and $\xi=0.3$. For $\xi=0.2$ the axis of winding
coincides with the centre of the pore, and we could classify this structure as analogous to
the $DH$ structure for smaller $R$. However, in line with the other structures for this radius
this could be seen as a merger of two equally sized clusters of 
oppositely tilted and connected stacked disc morphologies. Moreover, in the evolution
for $\xi=0.2$ (see figure \ref{fig9}e) we observe a 
fundamental mechanism: the orientation with respect to
the pore wall of the total structure, as well as the change of chirality of local structure, are mediated 
by melting and reconnection of helical sections. 
In this way, unfavorable orientation and winding in part of the structure (for instance, giving 
rise to holes or increased curvature) can be avoided. We conclude that all these structures
$-0.1 < \xi < 0.3$ are actually coexisting $L_{\perp}^{tilt}$ under two different angles. 
The helical connections between these two cluster structures are relatively stable in time,
and analogous of the double periodic array of saddle surfaces found
in lamellar forming thin films under an electric field.\cite{kyrylyuk06}
Apparently, they do not cost much energy to the system. 
This is confirmed by the free energy plot in Figure \ref{fig8}, were the free energies associated
with these different helical structures fall almost on a single line.\\
To see whether the elongational stress is the cause for this fundamental mechanism, we also considered 
the structural evolution in the absence of incommensurability along the pore, 
$\xi=0.3$ and $R=174 L_0$ (see figure \ref{fig9}f). 
The detailed analysis confirms that the fundamental mechanism
identified for formation of structure under elongational stress, i.e. reorientation on a higher structural 
level by helical connections, is
also instrumental for the formation of equilibrium $L_{\perp}$ structures. One can clearly see that
the tilted and interconnected discs that form in the initial stages
(1000 TMS) slowly transform into perfectly stacked discs or $L_{\perp}$ (25000 TMS). The mechanism is 
a moving front of helical connections, just like shown in Figure \ref{fig9}e.
This effect can also be observed for $R=2.21 L_0$ and $\xi=0.0$ or $0.2$ (not shown),
where the initial coexisting structure 
$L_{\perp}^{tilt}$/$L_{\perp}^{tilt}$ (TMS=10000) evolves into $L_{\perp}$ (TMS=20000). 
From these observation we conclude that coexisting stacked tilted discs with helical domains
are the preferred transient structure, independent of the elongational stress. 
The most probable cause is that interconnected tilted structures 
can much easier accommodate the necessary large scale 
structural rearrangements that are necessary for the formation of $L_{\perp}$.
Helical connections play an important role
in the removal of defects, and slowly change the tilt of the intermediate structure,
leading to the formation of equilibrium $L_{\perp}$ structures. 

\subsection{Discussion: comparison to other studies}

\subsubsection{Experiments}
Up to now, experimental groups have considered structure formation 
of (nearly) symmetric polystyrene-{\it block}-polybutadiene (PS-{\it b}-PB) \cite{shin04,xiang04,wu04,xiang05}
and polystyrene-{\it block}-poly(methyl methacrylate) (PS-{\it b}-PMMA)\cite{sun05}  
confined in nanopores. Almost all observed experimental microstructures belong to the class of
concentric cylinders. This does not come as a surprise, as for most diblock copolymers 
one of the blocks has a higher affinity for aluminum. 
We have previously concluded that, especially for the larger pore sizes considered, 
surface fields above a certain (relatively low) threshold value lead to 
so-called surface reconstructions: concentric cylinders. 
In the experimental procedure, the block copolymers were introduced 
into the alumina nanopores at elevated temperatures
via capillary action and subsequently cooled. After this procedure a weak base was
used to dissolve the alumina and produced the free standing rods
of block copolymer that were imaged. The thus obtained rods show several features: 
the cross-sections vary in shape (from ellipsoidal
to spherical) and the thickness of the rod varies substantially
along the rods (due to the roughness present in the alumina pores). As
a result, the actual pore radius can only determined approximately. Although 
some measurements of surface tension were performed,
exact numbers quantifying the affinity of $PS$, $PB$ and $PMMA$ for the
aluminium pore material are missing.\\
We conclude that the stacked-disc or toroidal-type structure in nearly
symmetric polystyrene-{\it block}-polybutadiene (PS-{\it b}-PB) block copolymer \cite{shin04} is
the best candidate for a true comparison between our calculations and experiments.
We therefore choose the simulation parameters as close as possible to the values for this experimental
system. The alternative structure was found at a temperature of $T=398.15 K$, above the
glass transition temperature \cite{xiang04,xiang05}, in a system with a molecular weight 
of $M=18400$ (the number of monomers $N \sim 250$). The experimentally 
measured repeat period $L_0=17.6 nm$, giving rise to an incommensurate 
pore radius $R/L_0 \sim 1.3$. We note that the effective FH parameter $\chi$ 
depends to some extend on the details of the experimental system. 
Using earlier measurements by the same group we find $\chi=0.058$ \cite{owens89} and 
obtain $\chi N \approx 14.5$. For a coarse-grained chain with
$N=22$ ($A_{10}B_{12}$) we therefore find an effective Flory-Huggins parameter
$\chi_{SB}=0.66$. Here we consider $\chi_{SB}=0.6$ instead,
and focus on the influence of reduced FH parameter in general 
(the calculations in the previous sections were carried out for $\chi_{SB}=0.8$). 
Although we use the bare FH parameters in
the calculations, the referring $\varepsilon^{0}_{SB}=1.5$ kJ/mol for the
choice of the temperature ($T=300 K$) in the standard input in DDFT.\\
The results of calculations for $\chi_{SB}=0.6$ and two radii $R$ (and varying $\xi$) are 
shown In figure \ref{fig13}a-b. We conclude that for the largest $R=19$ the effect 
of reduced $\chi$ is rather small. Comparing the results for $R=19$ in figure \ref{fig13} and 
$R=2.21 L_0$ in figure \ref{fig7} (for the same cylinder length $L_x$) we see that 
the phase boundaries do not been shift much.
For low and high values of $\xi$ concentric cylinders are found with an additional
layer when compared to the results for higher block-block interaction. 
This finding is well in agreement with the 2D SCF calculations of Li {\it et al} \cite{li06}
where a transition from $L_3$ to $L_2$ (notation of Ref. \cite{li06}) 
is found for $\chi N >13$. The tilt of the $L_{\perp}$ for small $\xi$ reflects the different 
bulk lamellar distance $L_0$ associated with the change in $\chi_{SB}$ and is due
to an extensional force. For the smallest $R=9$, concentric cylinders ($L_{||}$)
are found for $\xi \leq -0.1$ and $\xi \geq 0.5$, with the same number of layers
as for larger $\chi_{SB}$. For $0.1 \leq \xi \leq 0.3$, a perfect $L_{\perp}$ structure 
is obtained. Constructing the lines associated with different structures shows that
$\xi=0.0$ and $\xi=0.4$ are indeed located on intersections of different lines. Based
on the experimental knowledge that the $B$ block is preferentially found close to
the pore surface, we further concentrate on $\xi=0.4$. For $R=19$
the structure contains perforations, which will not be visible in the
experimental imaging procedure. This structure will appear as concentric cylinders
($L_{||}$). For $R=9$, the structure is a mixture of 
$L_{||}$ and $L_{\perp}$. The standard element are 
two opposite small lamellar patches oriented parallel to the pore surface. 
Along the pore, only the orientation
of the elements changes and neighbouring elements are orthogonal (see figure \ref{fig14}a). 
The fourfold connections (instead of three) between the elements are short 
and curved cylinders. This mixed structure has interesting features that somewhat 
resemble the experimental toroidal-type structure of Shin et al \cite{shin04}. However,
the structures still do not match in all details. Additionally, we varied the small pore radius 
for fixed $\xi=0.4$ (figure \ref{fig13}c). This leads to $L_{||}$
for $R=5, 8$ and $L_{\perp}$ for $R=6,7,10$ and $11$. In general, we conclude
that the decrease of $\chi N$ reliefs some of the stresses present in the stronger
phase segregated system, and leads to faster relaxation towards a stable structure.\\
Finally, the small scale of the experimental nanopores leads to constraints 
that are not effective in other systems. The results of the previous section show that 
the structure-type can be very sensitive to the pore radius. Small variations of the pore 
radius (due to roughness) along the pore at relatively larger length scales  
may therefore affect the structure formation to some extent. 
This is in particular the case in small nanopores, where 
slightly different pore radii will promote different structure-types
and connectivity issues (of different types and periodicity) may appear. 
To study this effect, we prepared a pore with two pore radii, $R=9$ and $10$ ($\xi=0.4$), 
equally distributed along the pore (the thickest part is located in the middle of the pore). 
Using $L_0=8.6$, the bulk characteristic lamellar spacing determined for the system
with higher $\chi_{SB}$, the relative pore sizes $R/L_0$ vary between $1.05$ and $1.16$.
Keeping in mind that $L_0$ will decrease for decreasing $\chi_{SB}$, 
this range matches the experimental value of $1.3$ reasonably well. 
Following the evolution of this structure (see figure \ref{fig14}b), we initially observe that simply
the stable structures for the individual pore radii are formed and connect 
($L_{||}$ for $R=10$ and the cage structure
for $R=9$). However, eventually the interface between the different structures
acts as a driving force for a transition to another structure. The final and stable structure
for the $S$ block is completely perpendicular, with stacked disc for $R=10$, but further away from this 
region the stacked discs become perforated and adapt the toroidal shape (see figure \ref{fig14}b,
second from left).
From a comparison between the experimental image
and the simulated 3D structure we observe that cross section of the $B$ block in the centre of the pore
(figure \ref{fig14}b, utter right)
looks very similar to figures 1C and D in Shin {\it et al}\cite{shin04}.
Since the structure for larger $R$ and the same parameters is $L_{||}$
we conclude that this computational system gives a very close match.

\subsubsection{Other numerical studies}
The differences in methodology and choices of the parameter 
range complicate the comparison to results of other computational studies in general. 
Although most of these studies where not aimed
at modelling experimental systems, we note that none of them have
reproduced the experimental toroidal-like structure.\cite{shin04} Moreover, the dynamics
in our method is aimed at following the experimental pathway. 
Here, we focus on a qualitative comparison to the two
most extensive MC studies for symmetric diblock copolymers in nanopores of three different radii:
Feng {\it et al} \cite{feng06a} and Chen {\it et al} \cite{chen06}.\\
In Feng {\it et al} \cite{feng06a} $\chi N$ is relatively low and 
stacked discs ($L_{\perp}$) are found for neutral pores, as expected. 
The region where $L_{\perp}$ is stable shrinks for increasing pore radius. For large surface
interactions, concentric cylinders ($L_{||}$) are found. 
The only alternative structures for intermediate surface fields
are $PL$ ($R/L_0 \approx 0.8$) and $L_{\perp,||}$ ($R/L_0 \approx 1.0$).
These results are fully in line with the findings in this article (see results section).
Moreover, for $R/L_0 \approx 1.0$ and an increased $\chi N$, also helices and double
helices were obtained for varying surface field strengths. We also find these structures
but for somewhat smaller $R$. Previously, we identified these structures as kinetically
trapped or stabilized by elongational forces. The particular surface field strength and
pore radius at which this effect plays an important role may be expected to depend on
the strength of the block-block interface, ie the value of $\chi N$. Moreover,
also the slight asymmetry of our own block copolymer may lead to a shift
of the parameter space where this effect is present.\\ 
In Chen {\it et al} \cite{chen06},
the value of $\chi N$ is considerably higher and the pore radii very small.
For small to zero surface fields several structures are found to be stable: stacked discs
$L_{\perp}$, helices $H$ and perforated (concentric) cylinders $PL$. Strangely enough
tilted stacked discs are absent, which may have to do with the absence of extensional
stresses for the considered pore length. Previously (previous section)
we found that for small radii and our choice of $\chi N$ that the free energy difference
between the helical and perforated lamellar structures can be small. 
Our results also suggested that the helical structure can be formed via the 
merging of perforations in the perforated structure, and subsequently that
$PL$ is an intermediate structure in the $D \rightarrow PL \rightarrow H$ evolution. 
The fact that here $PL$ is found for small surface fields is therefore not completely unexpected. 
For large surface fields, $L_{||}$ is stable.
The only true discrepancy between this study and ours is the stacked circle structure
found for $D=26$\cite{chen06} close to the stability region of the $L_{||}$ structure.
This structure, similar to $L_{||}$ but with short bounded instead
of an infinite cylinder in the pore centre, could originate from packing problems in combination with
a relatively large surface field. Although the bulk distance $L_0$ was not determined
explicitly, we can take $L_0=16$ from counting the number of stacked discs
in the cylinder, and find $R/L_0=0.81$. We refrain from further speculations about the
origin of this structure, and conclude that it is either an exotic structure or due to
the higher $\chi N$.\\
Finally, we briefly consider the results of a nanopore study that employed 
dissipative particle dynamics (DPD).
In Feng {\it et al} \cite{fengdpd06} a $A_5 B_5$ block copolymer system is simulated, 
both in neutral pores and pores with surface fields. The pore wall is not included as a geometrical
constraint, but as soft core DPD-beads with strong repulsion. The effect of
extensional forces is not considered. The authors find that 
$L_{\perp}$ always forms in neutral pores, except for the smallest pore radius considered 
($R=7$ \cite{fengdpd06}) where a double helix $DH$ 
was found. For pores with strong surface fields
only $L_{||}$ forms, except for the smallest radius where $L_{\perp}$ is obtained.
For surface field strength between these two extrema, several structures are
obtained and hopping between different states of the system is observed. Although
not mentioned as such, the structures in Figure 5 \cite{fengdpd06} show that
they can be arrested in alternative morphologies. 
In particular, helical structures were found to serve as intermediates. Although the parameter
range in this study is different and limited, it confirms 
our findings that the system can rather easily adapt to incommensurablity in
larger nanopores. Moreover, for the intermediate parameter range (small, nonzero
surface fields) there is a delicate balance between entropic and energetic 
contributions to the free energy, and structural transitions are mediated by helicity.
In our more systematic study we show that structures can sometimes 
be arrested in these transition (helical) states. 
Moreover, we showed that for very small pore radii 
perpendicular structures can be found for relatively large surface fields.

\section{Conclusions}
We have used a DDFT method to study the behavior of a slightly asymmetric
diblock copolymer $S_{10} B_{12}$
confined in a cylindrical nanopore. This block copolymer forms lamellae in the bulk, 
and was chosen to model the experimental system in Xiang et al \cite{xiang04}. 
We find many structures 
that were also identified in nanopore studies for completely 
symmetric diblock copolymers, but also a few new structures. However, almost all
exotic structures are metastable.
Our aim here is to identify
general rules for structure formation in nanopores.\\
We have focussed on the intermediate surface strength in the parameter space, where the effective 
surface field, determined by the difference of
the two block-surface interactions $\xi$, is relatively small. 
Outside this region, curved concentric cylinders (in analogy with the
thin film notation denoted by $L_{||}$) are stable based on energetic arguments.
The sign of the surface field determines
the block that is preferentially found to wet the pore surface: for 
positive $\xi$ the $B$ block is found close to the pore surface, 
for negative $\xi$ the $S$ block is preferential close to this surface.
Moreover, for a small and vanishing surface field 
the elastic chain deformations present in the concentric cylinders
are avoided by the formation of
a stacked disc structure (denoted as $L_{\perp}$).
The transition point from parallel structures ($L_{||}$) to 
perpendicular structures ($L_{\perp}$) is located
somewhere inside this region, and we find many alternative structures.
Analysis of the free energy and the determination of a phase
diagram of equilibrium structures shows that all of these structures
except $L_{||}$, $PL$ and $L_{\perp}$
are metastable, and we have considered
possible causes in detail.
For stronger surface fields, that in general favour surface reconstructions (parallel structures),
commensurability of the pore radius $R$ and the natural
lamellar domain distance $L_0$ is important to avoid large compression or extension
of the chains in parallel structures. For weaker surface fields, favouring
perpendicular structures, commensurability of the pore length $L_x$ and the natural
lamellar domain distance $L_0$ is important to avoid large extensional forces. 
The commensurability issues, together with sometimes extremely slow transition dynamics, were
found to cause the formation of
alternative structures.\\
For stronger surface fields, phase separation
is initiated by the presence of the interacting pore surface. As a result, the pathway
from the homogeneous melt to more stable alternative structures includes a first stage of
parallel structures or surface reconstructions, even when $L_{||}$ is not equilibrium.
Subsequently, when the driving force is relatively large,
holes are formed (for larger $R$, due to undulations 
that form connections between different layers) and connect: a helical structure
is formed at a larger scale. The most probable origin of this helicity is that helical structures 
can more easily accommodate structural reformations at a higher level by adapting their
spacing and pitch than defected $L_{\perp}$ structures. At the final stages, the
helical structures line up by defect movement: the formation and destruction 
of three-folded connections. We find that these helical structures cost few energy to the
system. Along this pathway, the structure can get kinetically trapped depending on the
frustration that it experiences. We find metastable $L_{||}$, structures with holes 
(denoted by $PL$, and similar to the mesh or catenoid structures in \cite{feng06a} and
\cite{chen06}, respectively), 
helices or double helices 
for decreasing surface field strength. 
In general, the transformation of the intermediate structures 
to other structures can be slowed down, or even be halted, by the presence of extensional 
forces due to incommensurability along the pore.\\
For weaker surface fields the influence of the confinement on the initial stages
is less substantial. As a result
the initial structure is very interconnected and defected. The mechanism for the 
removal of these defects, however, is very similar to the one described above: small
helical domains and three-fold connections play an important role. Here, the extensional
forces play a major role, and determines whether stacked discs, tilted stacked discs or
helical-like structures (which are actually clusters of tilted stacked discs under opposite angles,
connected via helical domains) are formed.\\
We have also considered the experimental system 
of Shin {\it et al}\cite{shin04}, in which a new toroidal-like structure was observed.
Our results suggest that for weaker surface fields and small pore radius the structures 
are very sensitive to pore radius variations. Moreover, these
pore radius variations give rise to interfaces between the stable structures
for constant pore radii that exist in parts of the pore. 
These interfaces act as the driving force for the transition to other structures.
For a certain surface field strength ($\xi=0.4$) and constant pore radius, 
we find a cage-like structure for a pore radius close to the experimental 
value, and perforated concentric cylinders (that appear as concentric cylinders in the
experimental imaging procedure) for large pore radii. 
Moreover, by including the roughness of the experimental pore into our 
calculation in the form of a step profile, we were able to
obtain a 3D structure that is very similar to the experimental structure.

\section{Acknowledgement}
We acknowledge support of NCF
(Stichting Nationale Computer Faciliteiten). We thank
Thomas Russell for pointing out the application of pores in nano-technology,
and Jan van Male for useful discussions.

\newpage
\section{Captions}
Table I: Radial distance between $S-B$ or block-surface interfaces 
(denoted by $D_i$, $\sum_i D_i=R$, counted from the pore surface) in parallel structures for $\xi=-1.0$ and
$\xi=0.7$. The position ${\bar R}$ of the interfaces is determined by the condition 
$\theta_S({\bar R})=\theta_B({\bar R})$.\\
\\  
Figure 1: Structural diagram with existing simulation results. 
Different symbols denote systems in different segregation regimes:
{\bf triangles up} - a $A_5 B_5$ system
with $N \chi \approx 15$ and {\bf triangles down}  
the same system
with $N \chi \approx 50$,\cite{feng06a}
{\bf circles} - a $A_8B_8$ system 
with $N \chi = 16$,\cite{Sevpore} 
{\bf squares} - a $A_{10} B_{10}$ system
with $N \chi \approx 100$.\cite{chen06} 
Open symbols represent stacked discs, closed symbols concentric cylinders
and red symbols alternative structures.
We have scaled the axes to make a comparison: the vertical axis shows the reduced pore radius $R/L_0$, 
with $R$ the pore diameter and $L_0$
the particular lamellar spacing in bulk, the horizontal
axis $|\varepsilon|$, the energetic interactions of A with the surface in the 
MC method. We have used the relation $\chi \approx 5 (\varepsilon/kT)$ \cite{freire03}
for the conversion of our energetic FH parameter.\\
\\
Figure 2: Nanopore structure diagram for varying pore radius (in $R/L_0$) and
surface field (in kJ/mol). Stable structures for each $R/L_0$ (the determination of which is
based on the free energies for constant pore length $L_x$, see in Figure \ref{fig8}) are denoted by a grey backgound.\\
\\
Figure 3: Representative 3-D nanopore structures for: a) 
$L_{||}$, b) $L_{\perp}$, c) $L_{\perp}^{tilt}$, 
d) $L_{\perp,||}$, e) $PL$, f) $H$, g) $DH$, h) $GH$.
The corresponding locations in the structure diagram are:
a) (1.74,-0.3) and (0.93,0.7), b) (1.74, 0.1) and (0.70,0.2), c)
(0.93,0.1) and (1.05,0.1), d) (0.93, -0.1), e) (0.70,-1.0) and (0.70,0.7),
f) (0.70, -0.1) and (0.70,-0.2), g) (0.81,-0.5), and h) (0.93,0.3), (1.05,0.5) and (1.28,0.4).
Isodensity surfaces of the $S$ component for mean value are shown.\\
\\
Figure 4: Representative 3-D nanopore structures for: 
a) $dL_{||}$, b) $dL_{\perp}$, c) $dL_{\perp}^{tilt}$, d) $L_{\perp}^{tilt}/L_{\perp}^{tilt}$, 
e) $dL_{\perp,||}$, f) $dPL$, g) $dDH$, h) $dGH$.
The corresponding locations in the structure diagram are:
a) (0.93,-0.5), b) (2.21, 0.3) and (1.74,0.4), c)
(2.21,-0.1) and (1.05,-0.1), d) (1.28, 0.1), e) (0.93,0.0),
f) (1.05,-0.3) and (1.74,-0.2), g) (0.81,-0.1) and (0.81,-0.3), and h) (1.05,0.3) and (1.05,0.4).
Isodensity surfaces of the $S$ component for mean value are shown.\\
\\
Figure 5: Free energy (vertical axis) versus surface field strength $\xi$ (horizontal axis) for the reduced
radii considered: a) 0.58, b) 0.70, c) 0.81, d) 0.93, e) 1.05, f) 1.28, g) 1.74 and h) 2.21.
The free energies associated with the structures in Figs \ref{fig5}
are denoted by open circles (for $L_x=32$) and open squares (for $L_x=34$ or $36$).
Similar structures are connected by straight lines to guide the eye.
The open squares in a)-f) are associated with extra calculations (in all cases $L_{\perp}$)
for different $L_x$: a) 34, b)-e) 36 and f) 34. In c) the free energy associated with
the $DH$ structure ($\xi=-0.5$) is calculated for other $\xi$ (open diamonds). \\
\\
Figure 6: Nanopore phase diagram for varying pore radius (in $R/L_0$) and
surface field (in kJ/mol). Stable structures for each $R/L_0$ were determined
using all free energy data in Figure \ref{fig8}, irrespective of the pore lengths $L_x$.
The absence of $L_{||}$ for $\xi<0$ ($R=0.58, 0.70$) and $\xi>0$ ($R=0.70$)
does not allow for the analysis of the boundaries of the $PL$ region.\\
\\
Figure 7: Kinetics of nanopore structure formation for selected pore radii 
and surface field strength (in kJ/mol): a) $R/L_0=0.70$, $\xi=-0.4$, b)
$R/L_0=0.70$, $\xi=0.1$, c) $R/L_0=0.81$, $\xi=-0.5$, d) $R/L_0=0.93$, $\xi=-0.1$,
e) $R/L_0=1.28$, $\xi=0.2$, f) $R/L_0=1.74$, $\xi=0.3$. The images show isosurfaces 
for the $S$-block, for isosurface value $\bar{\theta}_S$. The numbers indicate
the dimensionless time steps TMS.\\
\\
Figure 8: The effect of extensional stress.
Kinetics of nanopore structure formation for $R/L_0=0.70$, $\xi=-0.5$ and varying
pore length $L_x$. a) $L_x=32$ (stable structure: $H$) and b) $L_x=36$
(stable structure: $L_{\perp}$). The images show isosurfaces 
for the $S$-block, for isosurface value $\bar{\theta}_S$. The numbers indicate
the dimensionless time steps TMS.\\
\\
Figure 9: The effect of extensional stress for the double helical structure
($R/L_0=0.81$, $\xi=-0.5$). Upon a variation of the
pore length $L_x$ a number of metastable structure classes (denoted by $DHi$ ($i=1-3$))
were identified.
The images show isosurfaces 
for the $S$-block, for isosurface value $\bar{\theta}_S$. Numbers indicate
the pore length $L_x$ (in grid units).\\
\\
Figure 10: Evolution of the free energy associated with the structures in Figure
\ref{fig11}. Numbers indicate the pore length $L_x$.\\
\\
Figure 11: Nanopore structures for $S_{10} B_{12}$ ($\chi=0.6$, $L_x=32$), matching
the experimental system of Shin {\it et al} \cite{shin04}, a) for pore radii 
$R=19$ ($L_x=36$), b) $R=9$ ($L_x=32$) and surface field $-1.0 \leq \xi \leq 0.7$ (in kJ/mol).
Parallel structures for surface fields higher than the threshold surface
field strength are not shown. 
The structures in c) are for constant surface field ($\xi=0.4$) 
and
pore radius varying between $5$ and $11$ ($L_x=32$).
For each radius, $R$, structures from two
different viewing angles are shown. 
The images show isosurfaces for the $S$-block, for isosurface value $\bar{\theta}_S$.\\
\\
Figure 12: a) High isodensity image of the cage-like structure for $R=9$ and $\xi=0.4$. b)
Structures for the system of Figure \ref{fig11} in a rough pore and $\xi=0.4$. 
The pore radius varies between $R=10$ (center region) and $R=9$ (outer regions).
From left to right: isosurface for the $S$-block (isosurface value $\bar{\theta}_S$) after
$1000$ TMS, the same for $20000$ TMS, isosurface of the $B$ block 
(isosurface value $\bar{\theta}_B$) after $20000$ TMS, crop of the isosurface at $20000$ TMS combined with
an orthogonal slice of the $B$ field in the centre of the pore.
\\
\newpage

\begin{table}
\begin{tabular}{||c|c|c|c|c|c|c||c|c|c|c|c|c||}
\hline
$R/L_0$&structure&$D_1$&$D_2$&$D_3$&$D_4$&$D_5$&
structure&$D_1$&$D_2$&$D_3$&$D_4$&$D_5$\\
\hline
0.58&$PL(SB)$&0.5&4.5&&&
&$L_{||}(BS)$&1.3&3.7&&&\\\hline
0.70&$PL(SB)$&1.6&4.4&&&
&$PL(BS)$&1.8&4.2&&&\\\hline
1.74&$L_{||}(SBSB)$&1.5&4.9&4.1&4.5&
&$L_{||}(BSBS)$&1.8&4.2&4.9&4.1&\\\hline
0.81&$L_{||}(SBS)$&1.2&3.4&2.4&&
&$L_{||}(BSB)$&1.4&3.1&2.5&&\\\hline
0.93&$dL_{||}(SBS)$&1.4&3.6&3.0&&
&$L_{||}(BSB)$&1.4&3.3&3.3&&\\\hline
1.05&$L_{||}(SBS)$&1.3&4.2&3.5&&
&$L_{||}(BSB)$&1.5&3.7&3.8&&\\\hline
1.28&$L_{||}(SBS)$&1.5&5.0&4.5&&
&$L_{||}(BSB)$&1.8&4.4&4.8&&\\\hline
2.21&$L_{||}(SBSBS)$&1.5&4.8&4.1&4.8&3.8
&$L_{||}(BSBSB)$&1.8&4.0&4.9&4.0&4.3\\
\hline
\end{tabular}
\label{table1}
\end{table}

\newpage

\begin{onecolumn}

\begin{figure}
\setlength{\unitlength}{1cm}
\begin{picture}(10,18)
\put(3.6,0){\epsfig{file=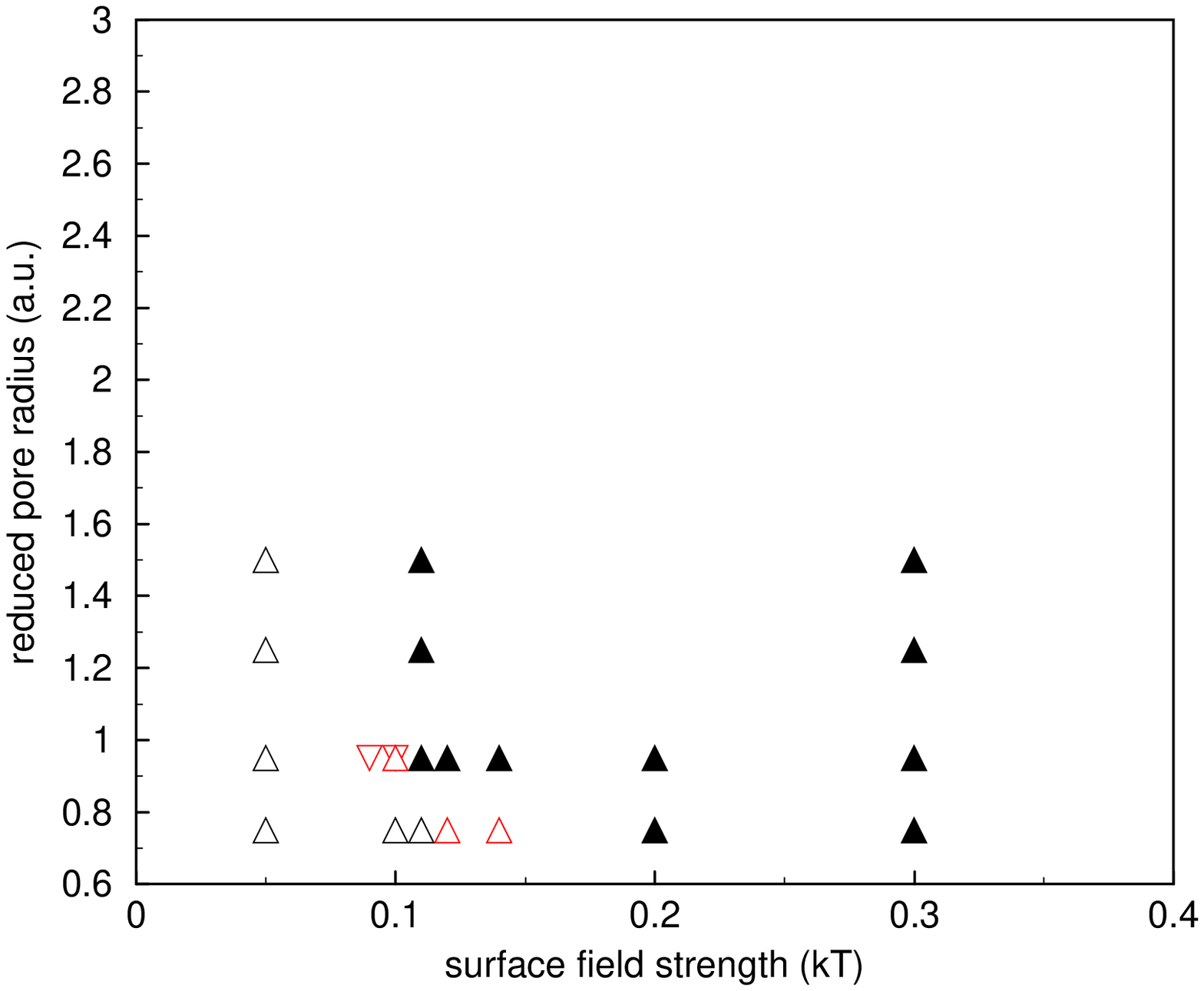,height=6cm}}
\put(3.6,6){\epsfig{file=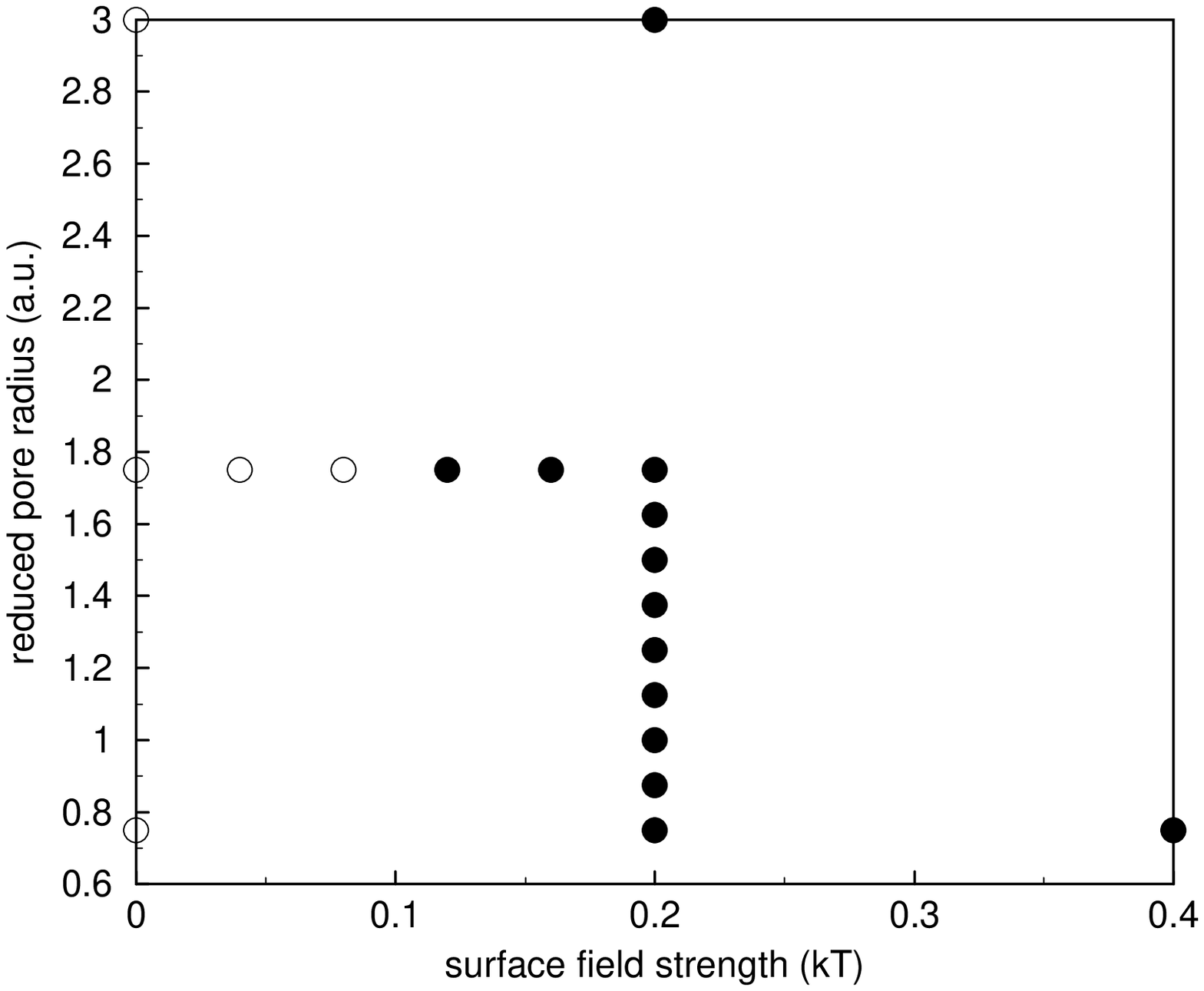,height=6cm}}
\put(3.6,12){\epsfig{file=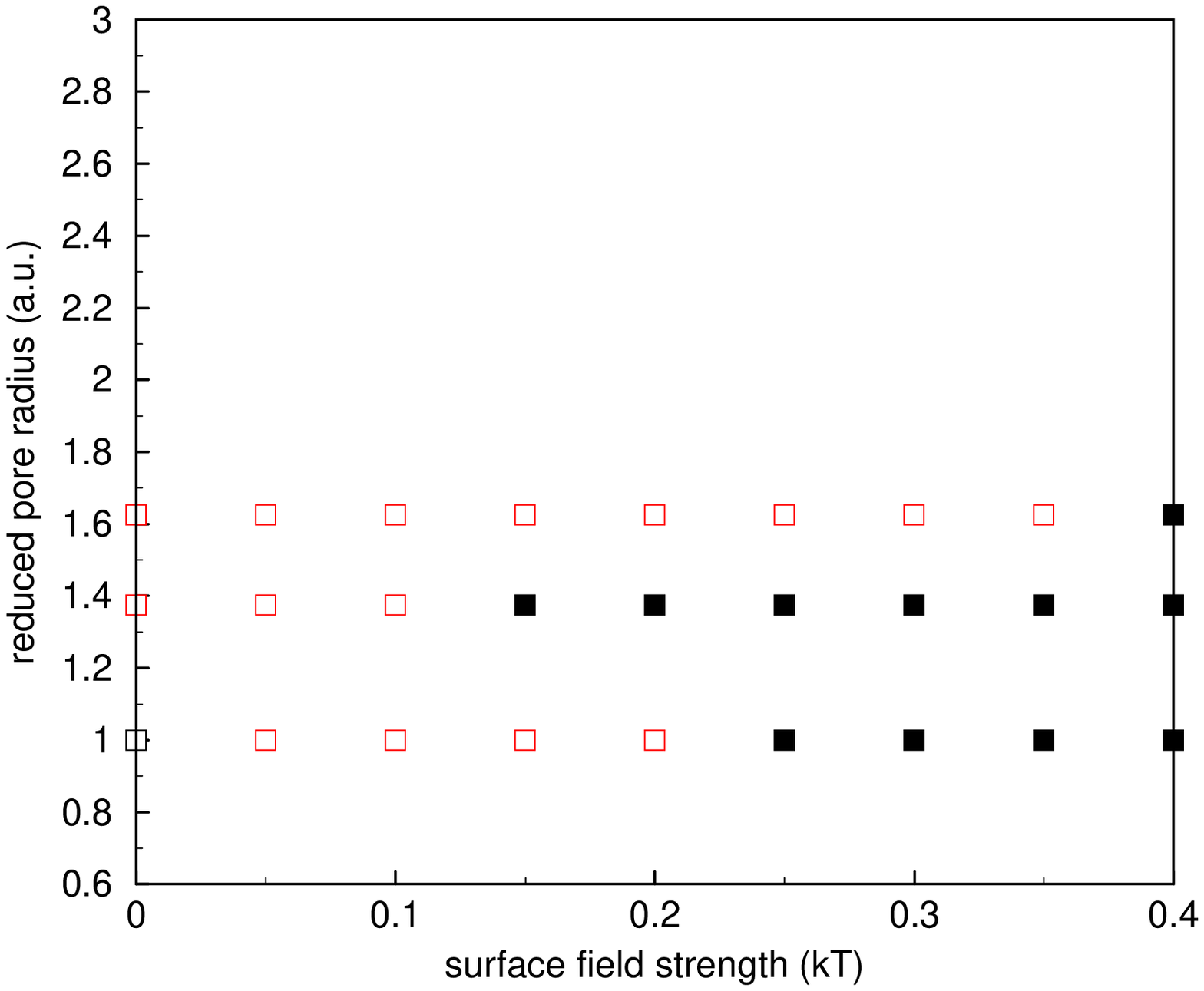,height=6cm}}
\end{picture}
\caption{}
\label{fig1}
\end{figure}

\newpage

\begin{figure}
\setlength{\unitlength}{1cm}
\begin{picture}(15,15)
\put(0,0){\epsfig{file=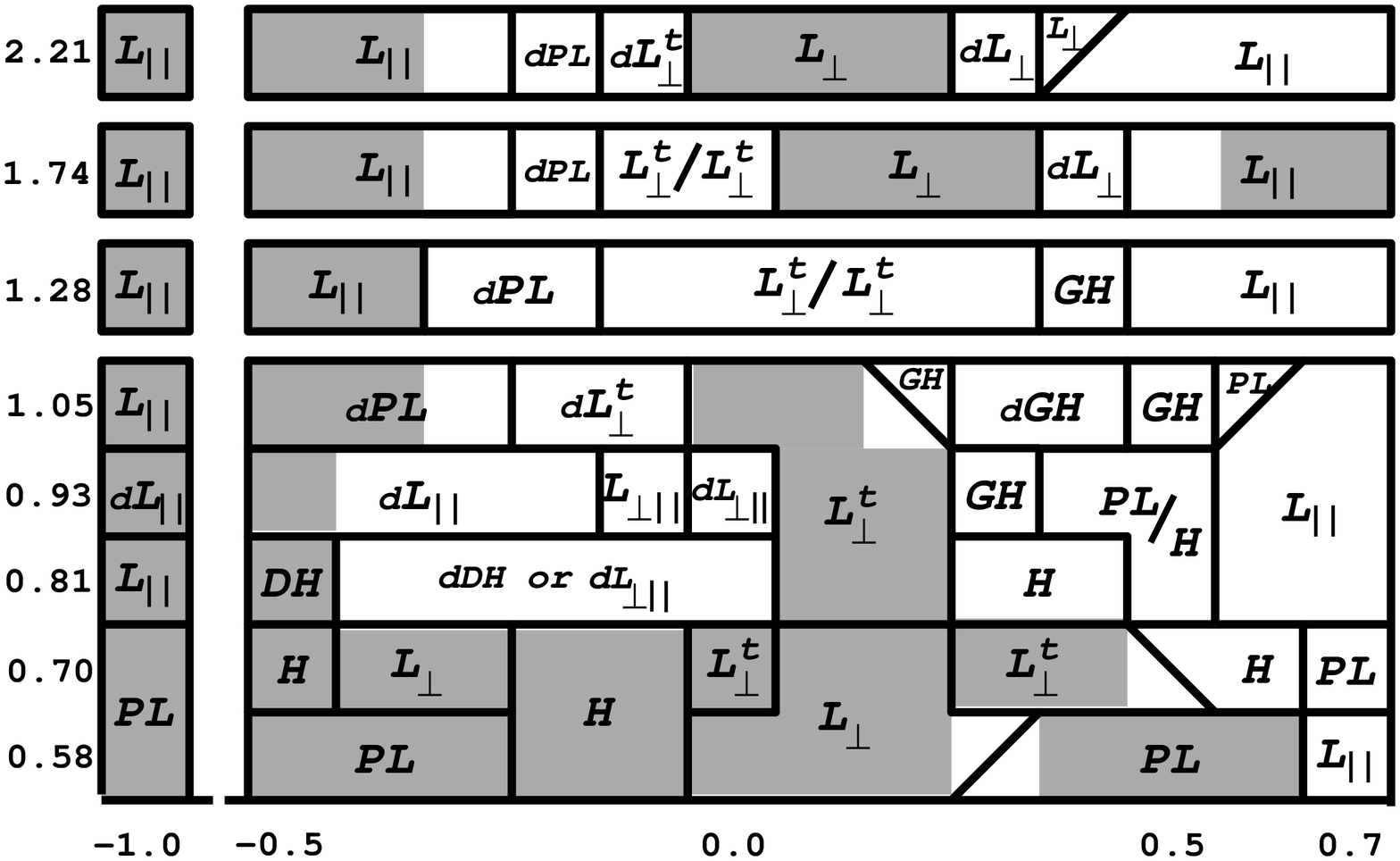,width=15cm}}
\end{picture}
\caption{}
\label{fig5}
\end{figure}

\newpage

\begin{figure}
\setlength{\unitlength}{1cm}
\begin{picture}(10,12)
\put(2.5,0){\epsfig{file=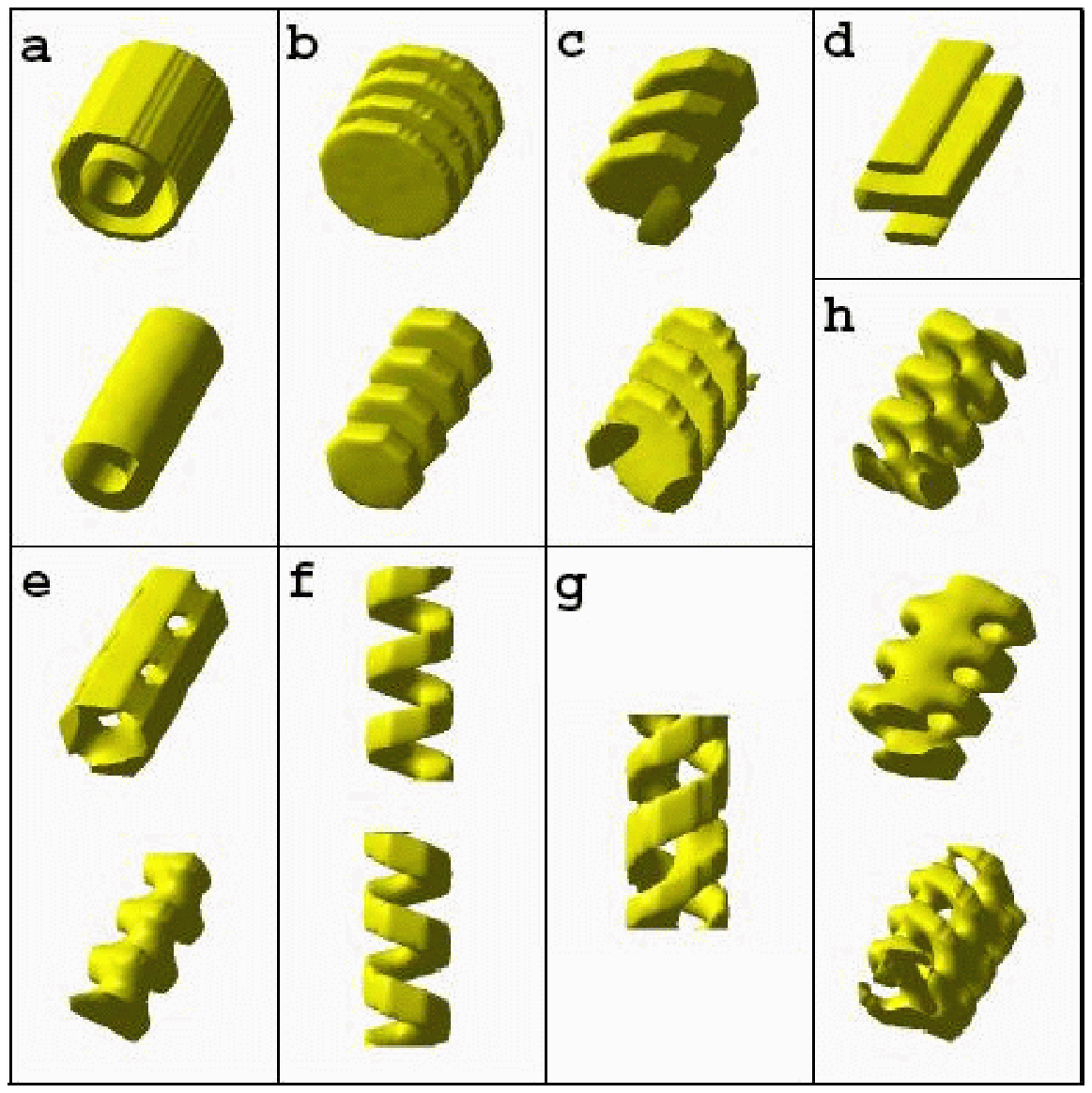,height=8cm}}
\end{picture}
\caption{}
\label{fig6}
\end{figure}

\newpage

\begin{figure}
\setlength{\unitlength}{1cm}
\begin{picture}(10,12)
\put(2.5,0){\epsfig{file=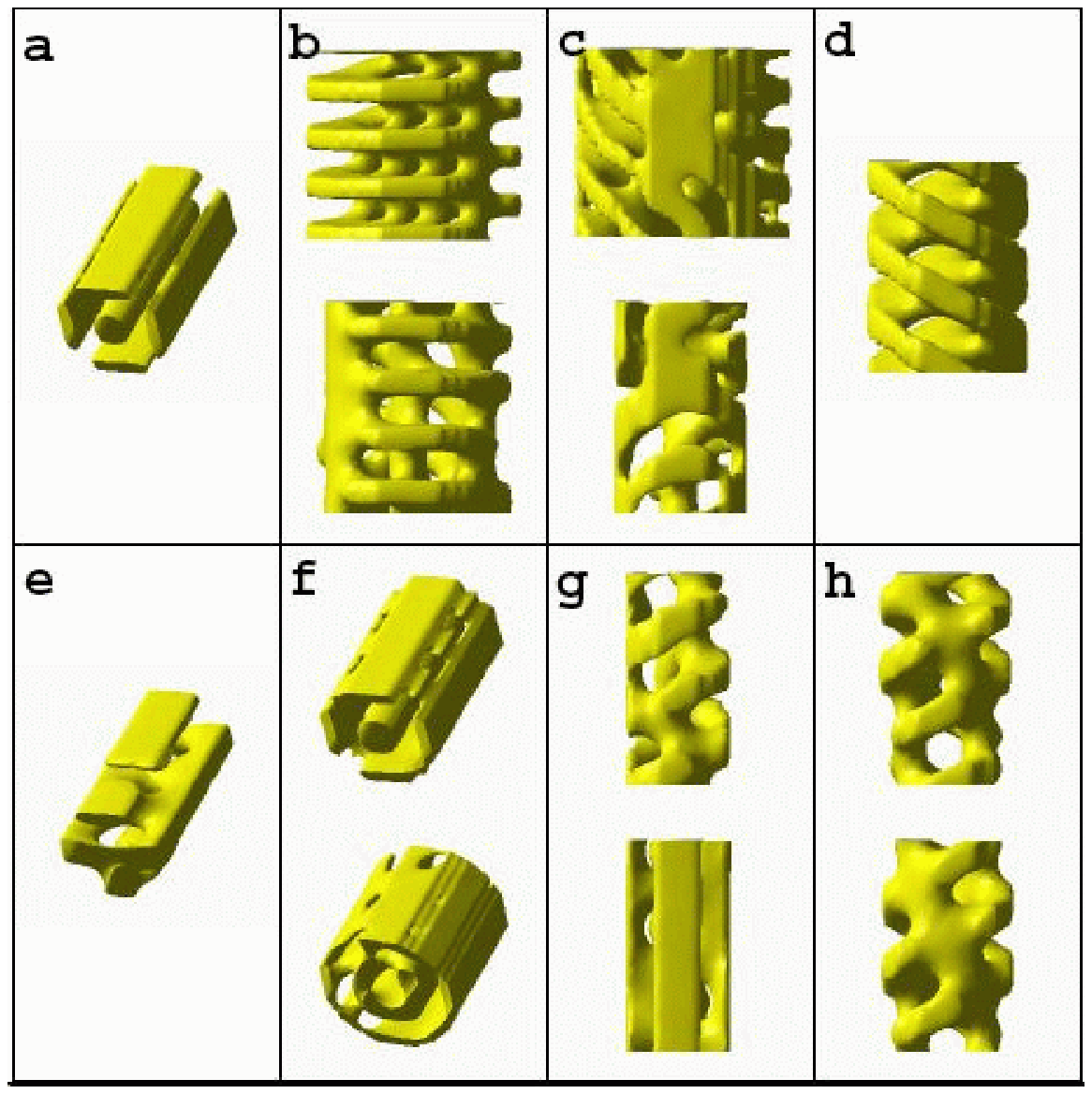,height=8cm}}
\end{picture}
\caption{}
\label{fig7}
\end{figure}

\newpage

\begin{figure}
\setlength{\unitlength}{1cm}
\begin{picture}(14,20)
\put(0.0,15){\epsfig{file=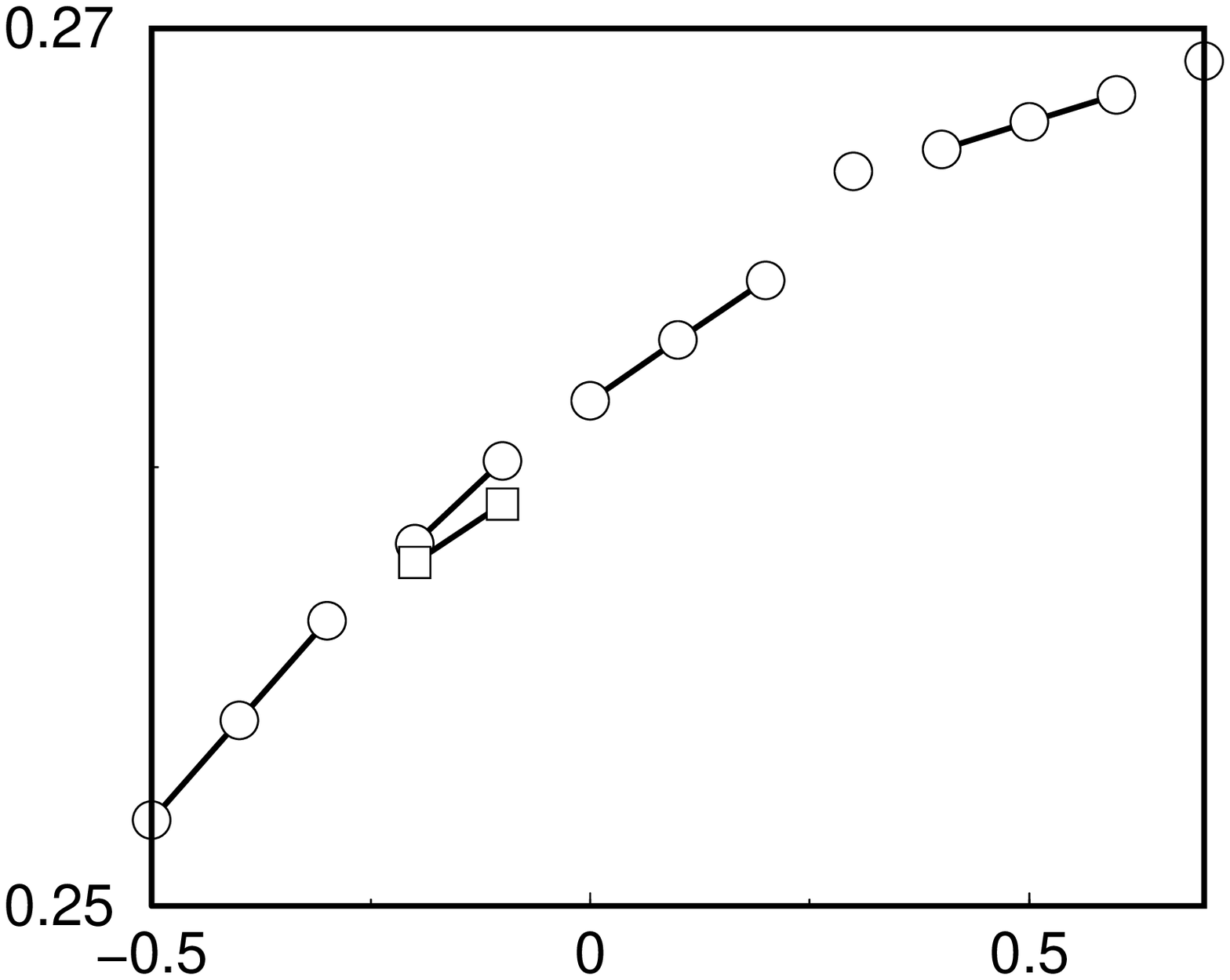,height=4.8cm}}
\put(7.0,15){\epsfig{file=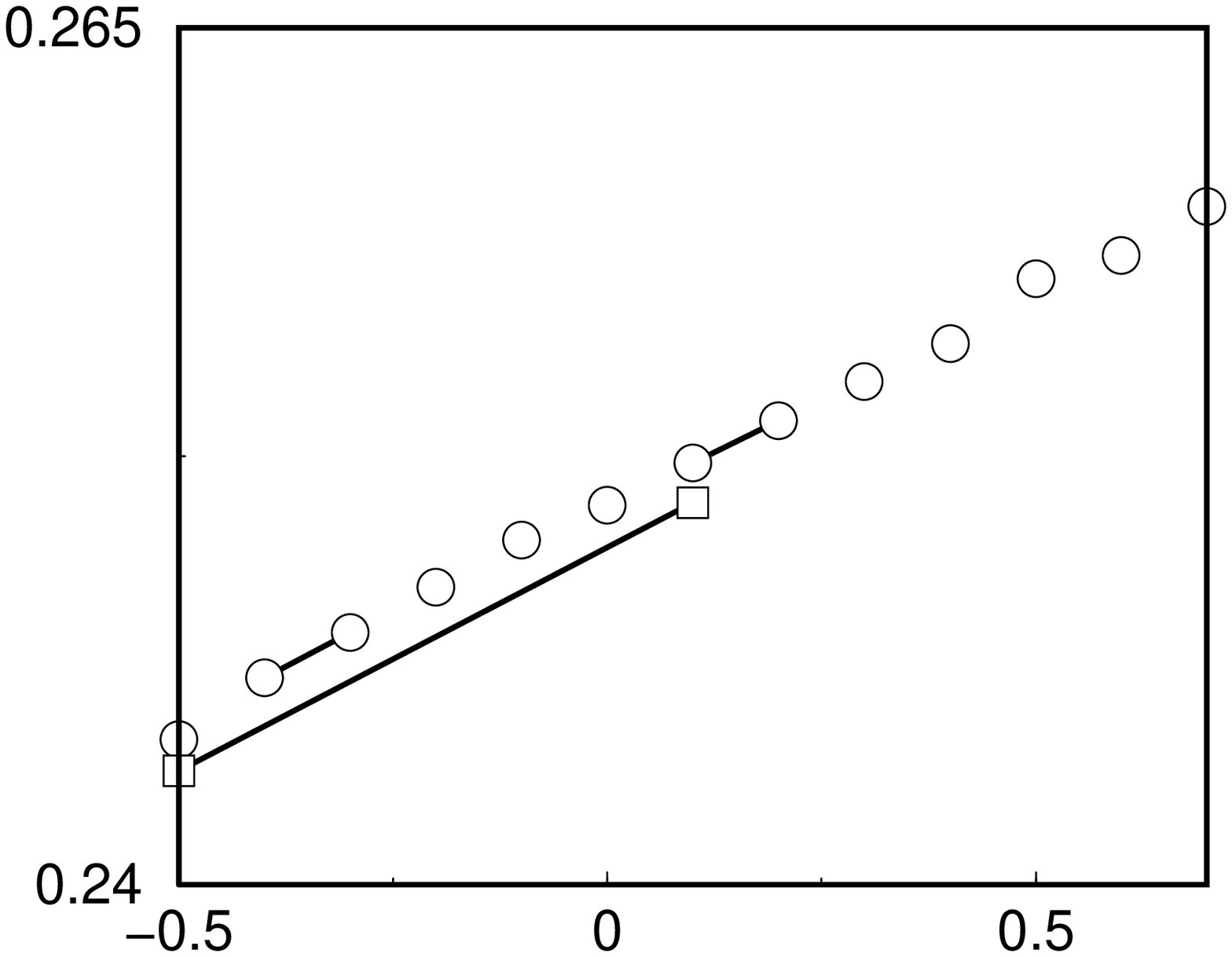,height=4.8cm}}
\put(-0.2,10){\epsfig{file=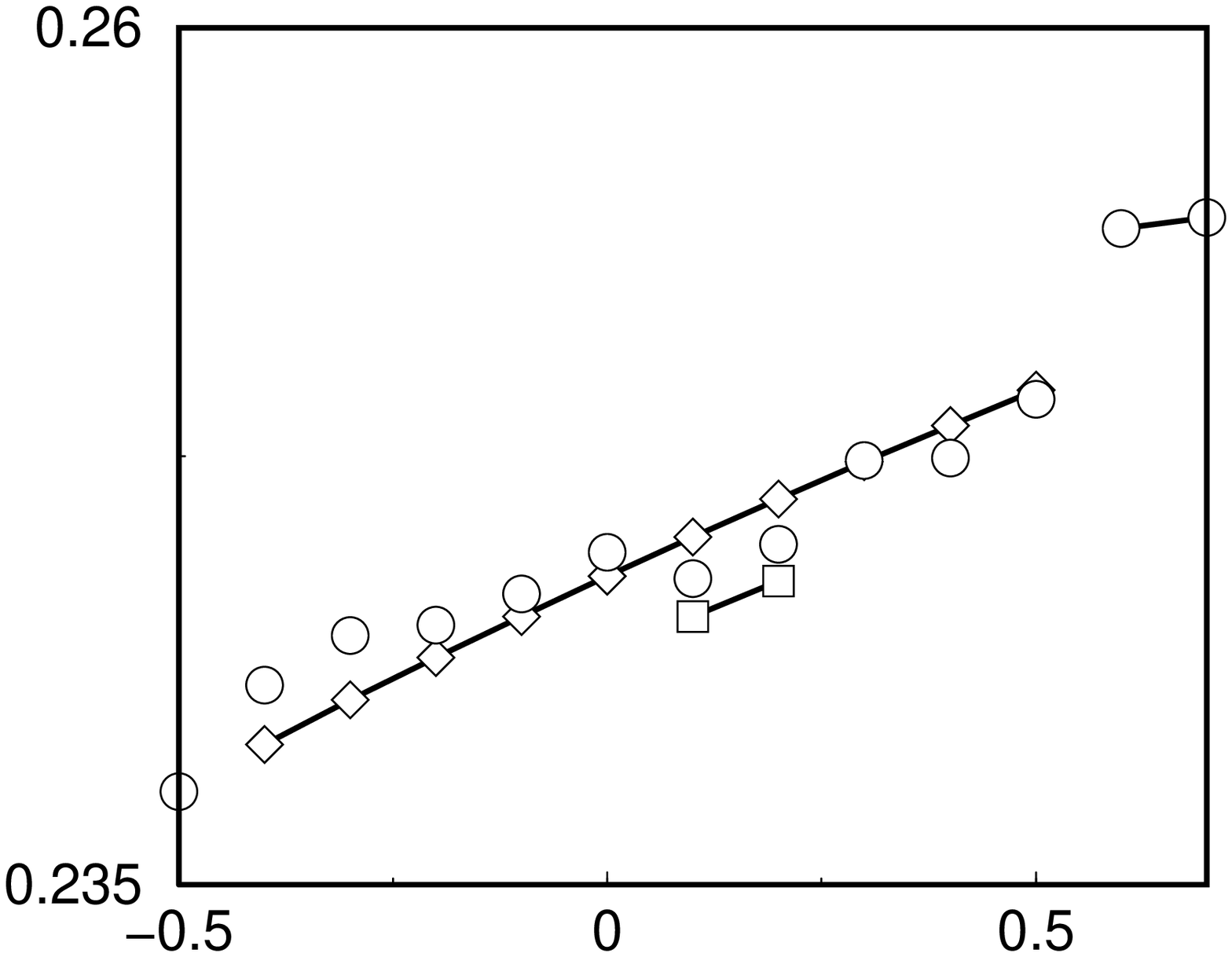,height=4.8cm}}
\put(7.0,10){\epsfig{file=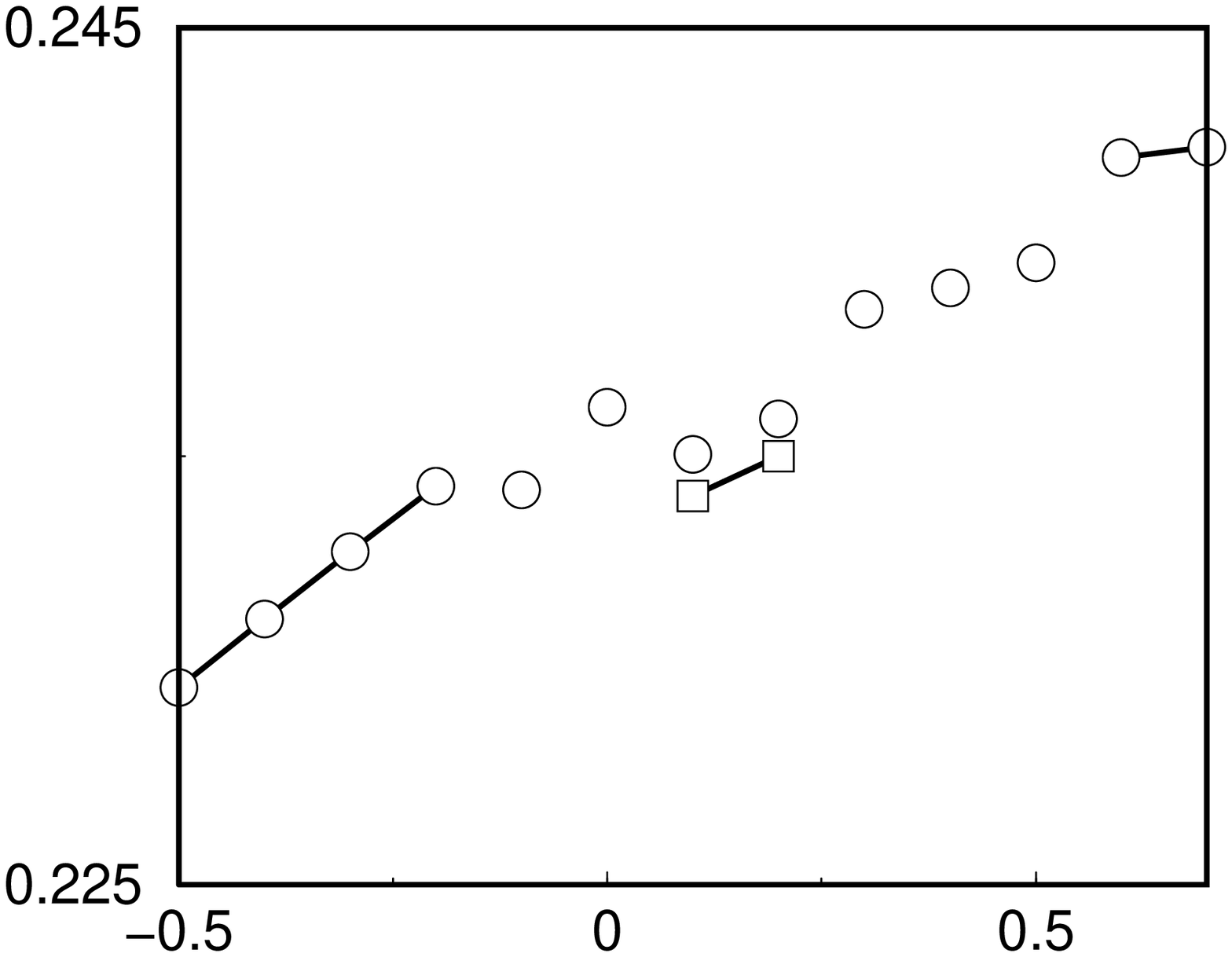,height=4.8cm}}
\put(0.0,5.0){\epsfig{file=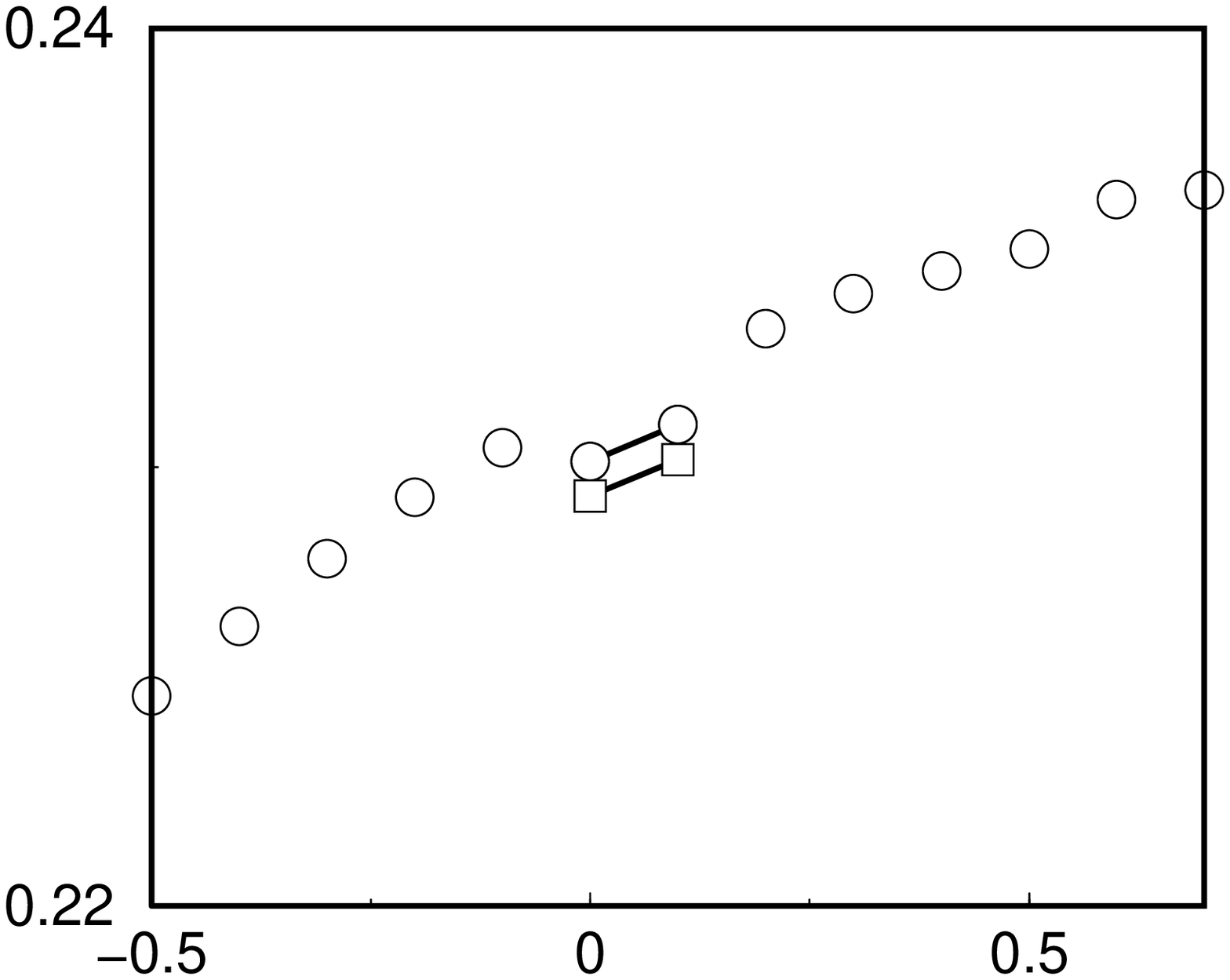,height=4.8cm}}
\put(7.0,5.0){\epsfig{file=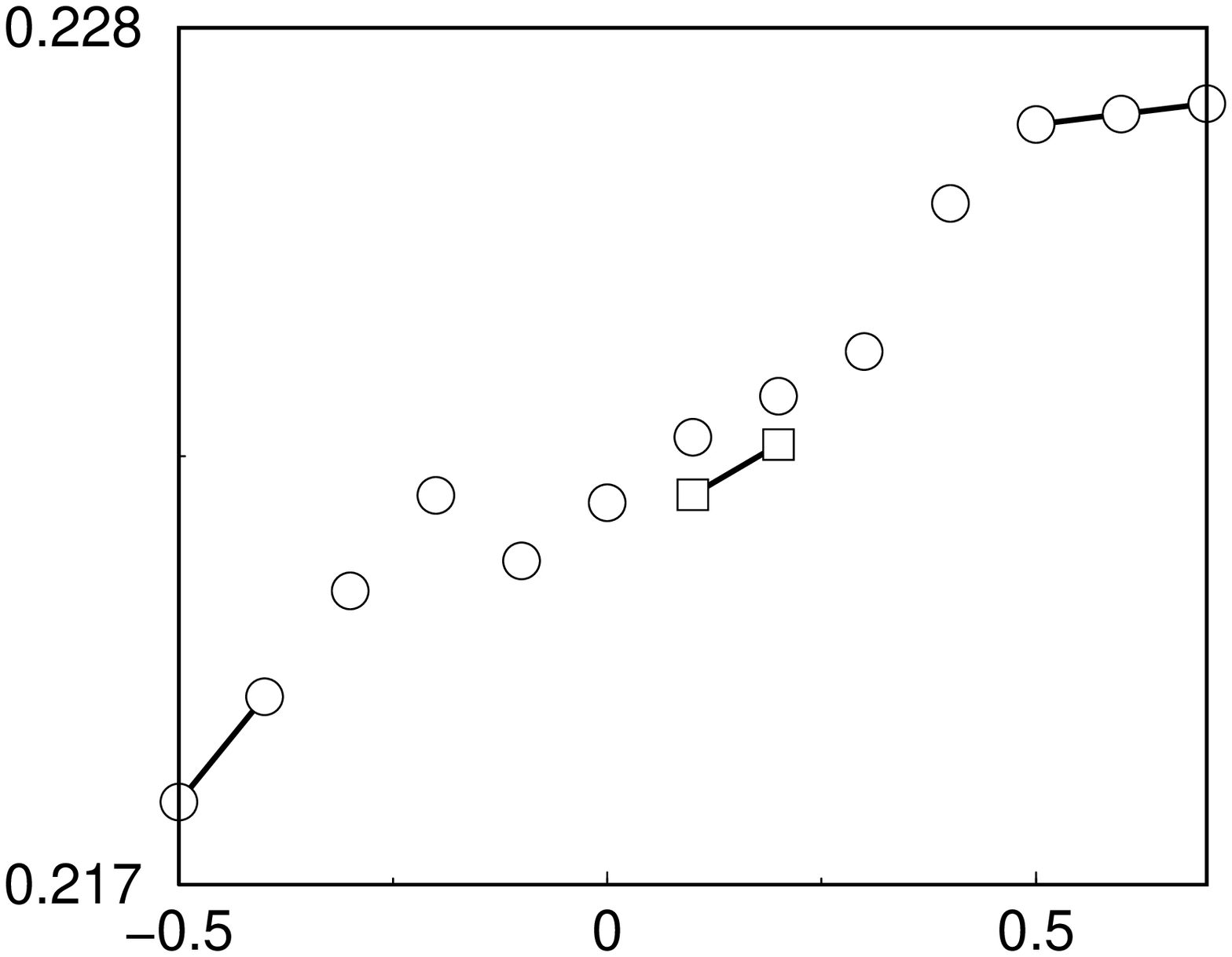,height=4.8cm}}
\put(-0.2,0){\epsfig{file=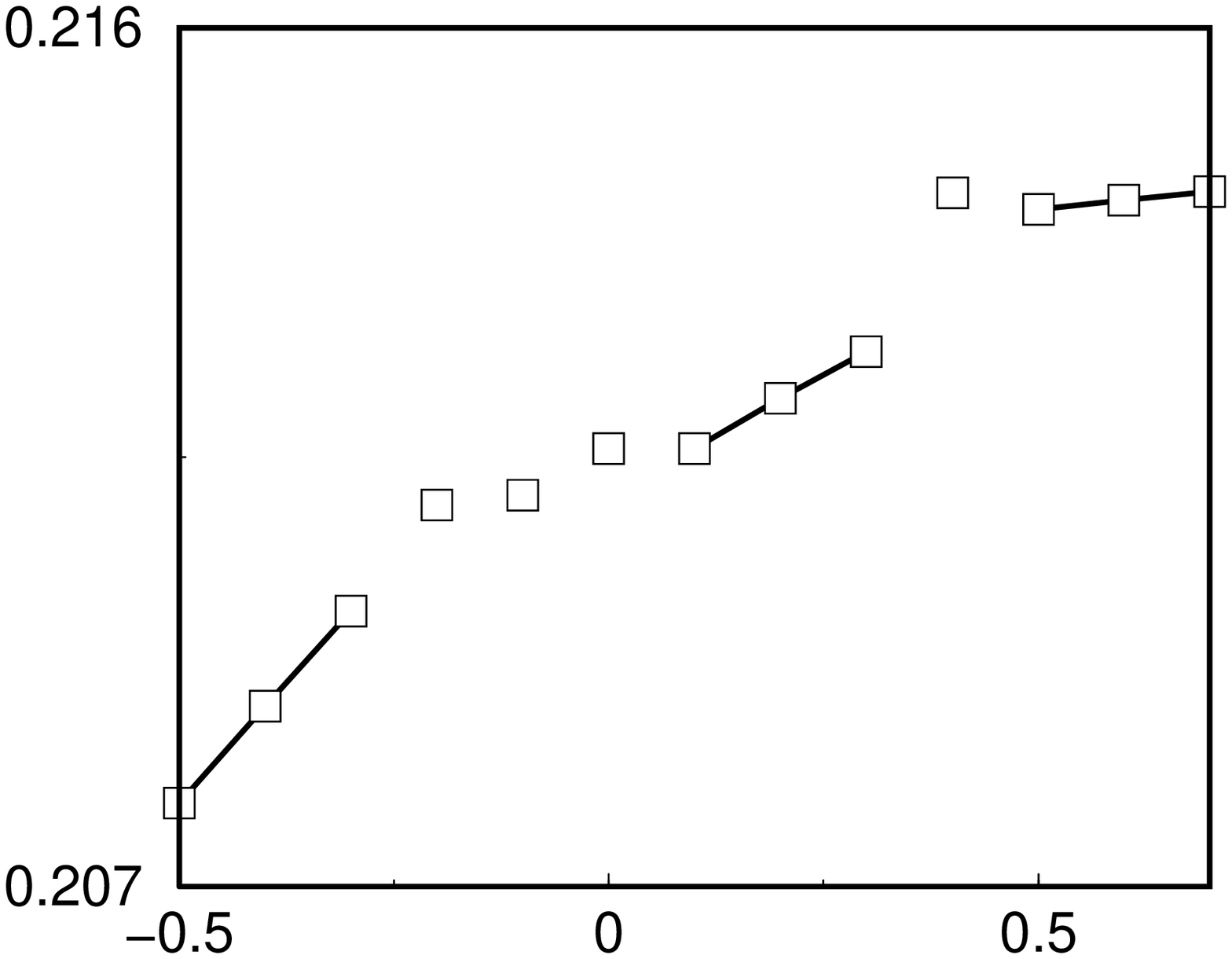,height=4.8cm}}
\put(7.0,0){\epsfig{file=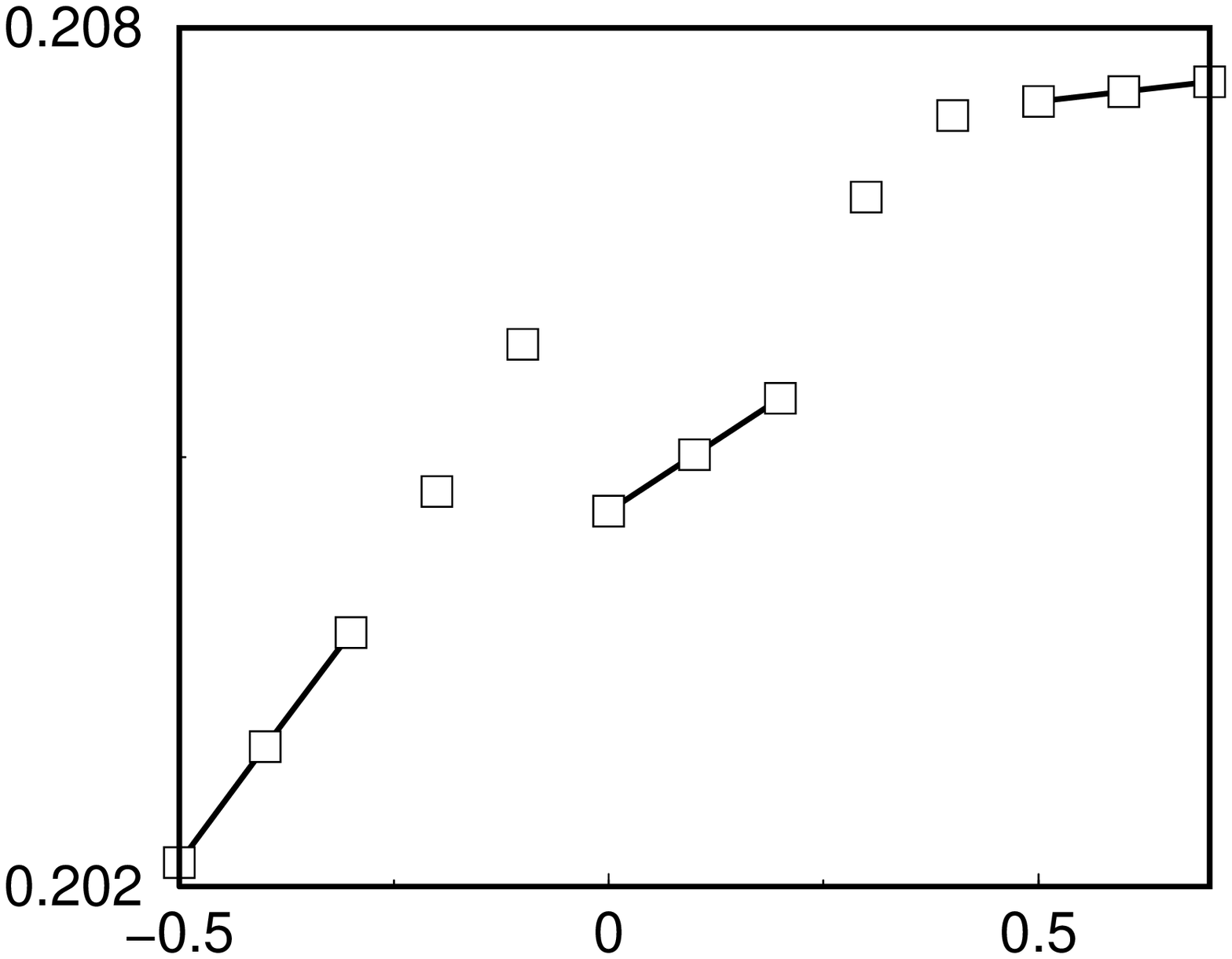,height=4.8cm}}
\put(0.0,19.0){a)}
\put(7.0,19.0){b)}
\put(0.0,14.0){c)}
\put(7.0,14.0){d)}
\put(0.0,9.0){e)}
\put(7.0,9.0){f)}
\put(0.0,4.0){g)}
\put(7.0,4.0){h)}
\end{picture}
\caption{}
\label{fig8}
\end{figure}

\newpage

\begin{figure}
\setlength{\unitlength}{1cm}
\begin{picture}(10,12)
\put(0,0){\epsfig{file=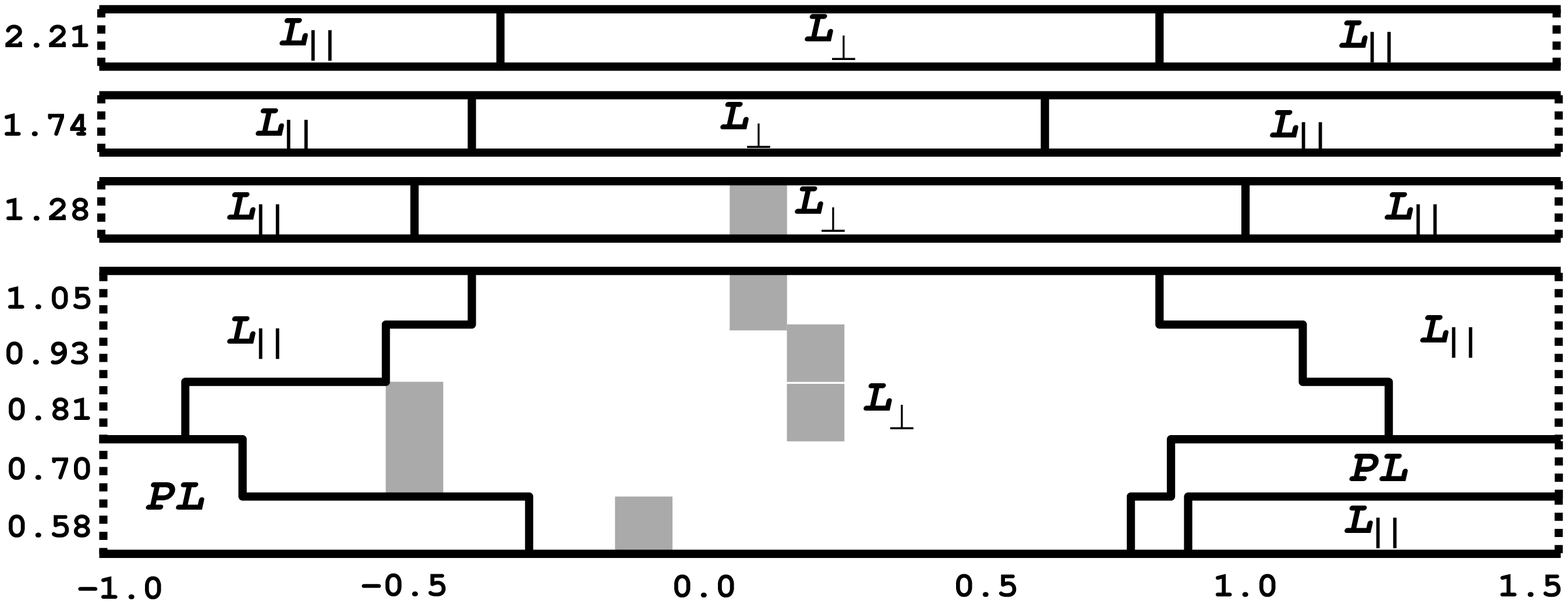,width=12cm}}
\end{picture}
\caption{}
\label{newfigure}
\end{figure}

\newpage
\begin{figure}
\setlength{\unitlength}{1cm}
\begin{picture}(14,14)
\put(2,4.5){\epsfig{file=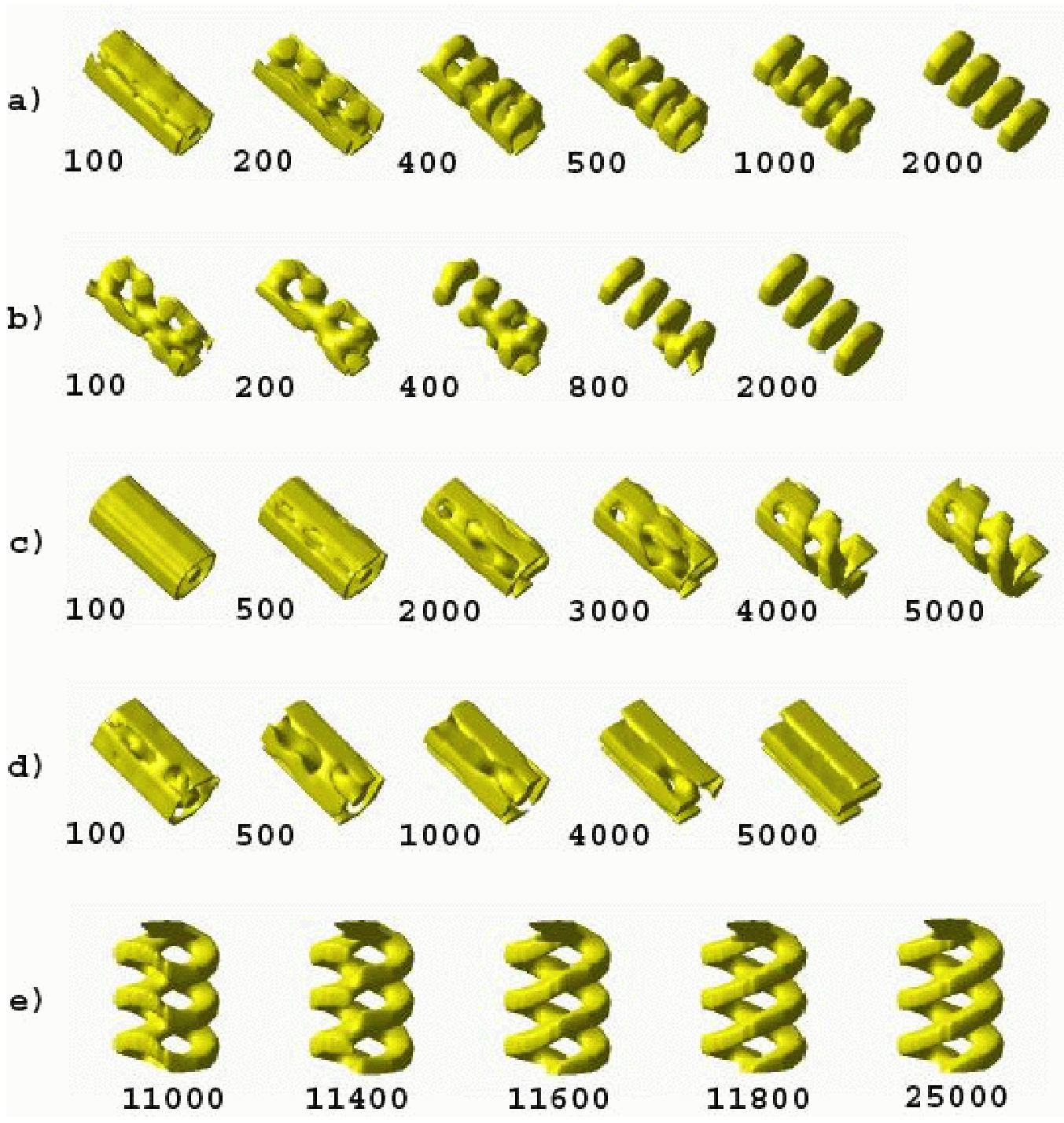,width=10cm}}
\put(2,0){\epsfig{file=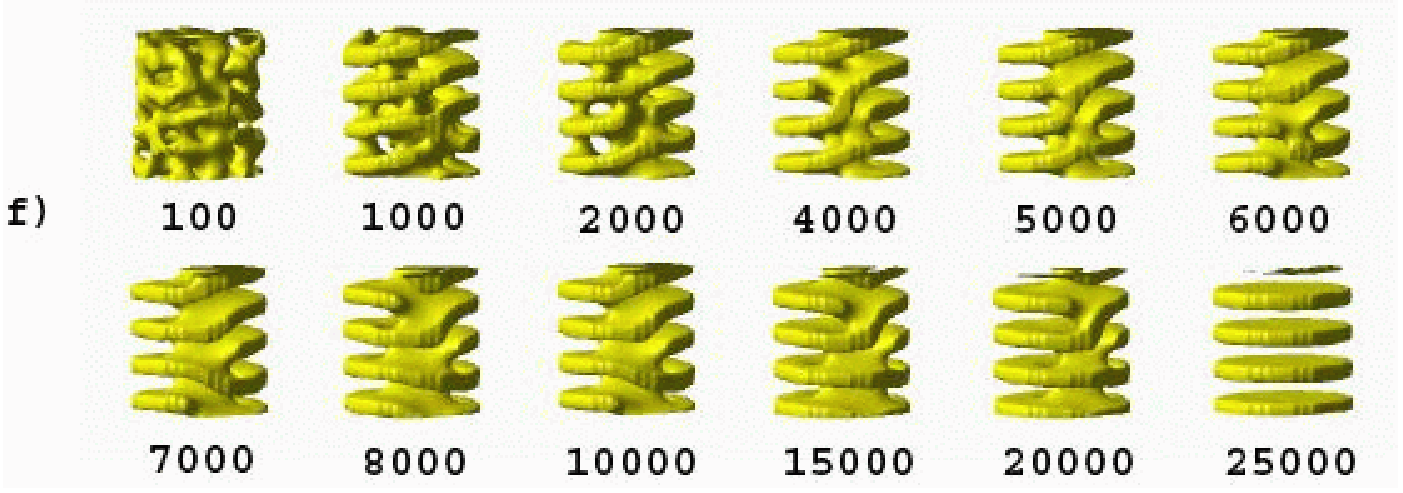,width=11cm}}
\end{picture}
\caption{}
\label{fig9}
\end{figure}

\newpage

\begin{figure}
\setlength{\unitlength}{1cm}
\begin{picture}(14,6)
\put(2,0){\epsfig{file=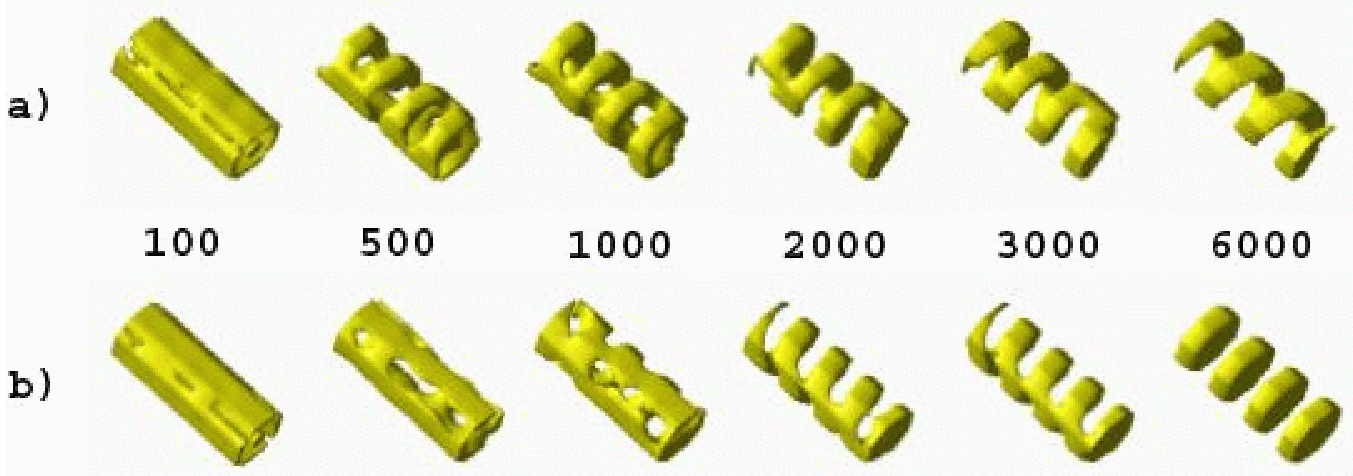,width=10cm}}
\end{picture}
\caption{}
\label{fig10}
\end{figure}

\newpage

\begin{figure}
\setlength{\unitlength}{1cm}
\begin{picture}(14,6)
\put(0,0){\epsfig{file=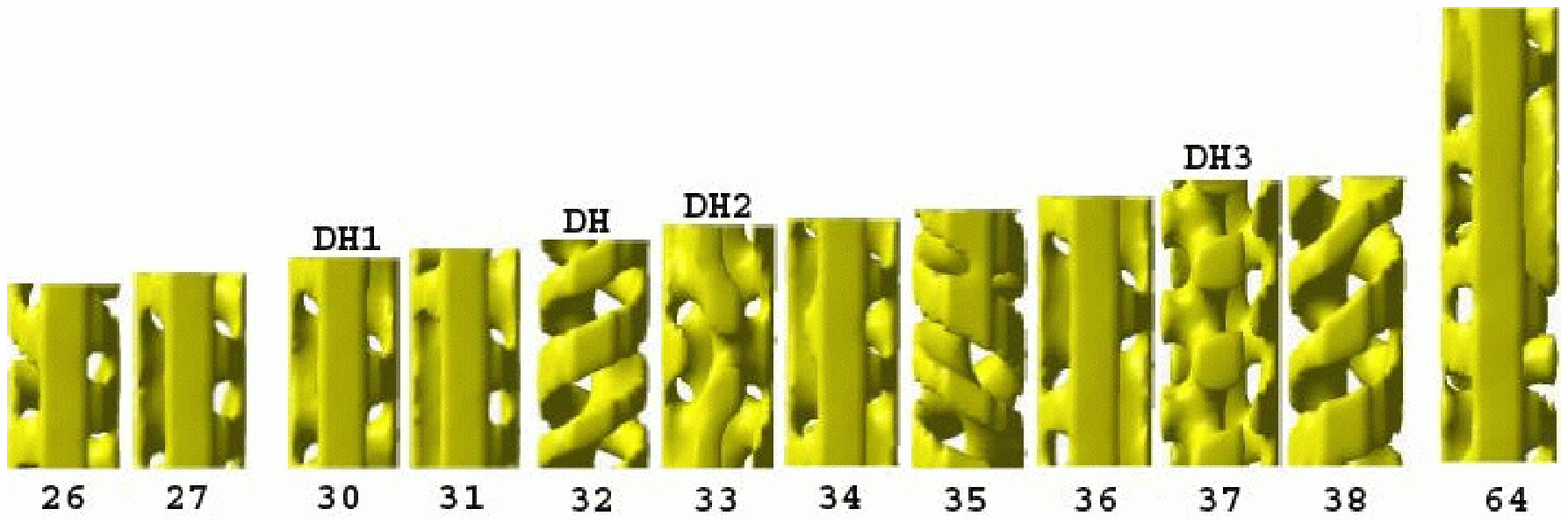,width=14cm}}
\end{picture}
\caption{}
\label{fig11}
\end{figure}

\newpage

\begin{figure}
\setlength{\unitlength}{1cm}
\begin{picture}(10,10)
\put(2,0){\epsfig{file=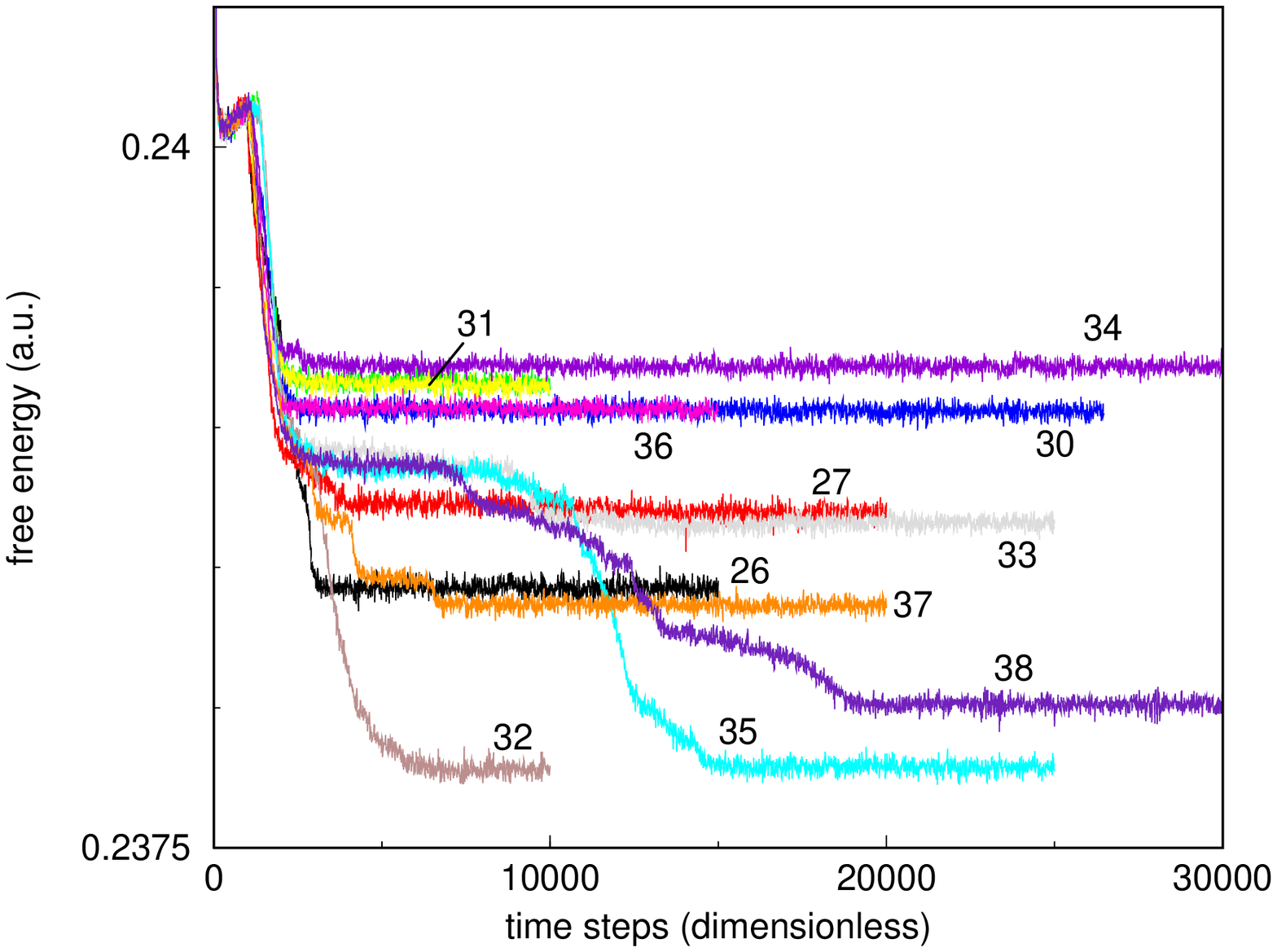,height=8cm}}
\end{picture}
\caption{}
\label{fig12}
\end{figure}

\newpage

\begin{figure}
\setlength{\unitlength}{1cm}
\begin{picture}(14,12)
\put(2,8){\epsfig{file=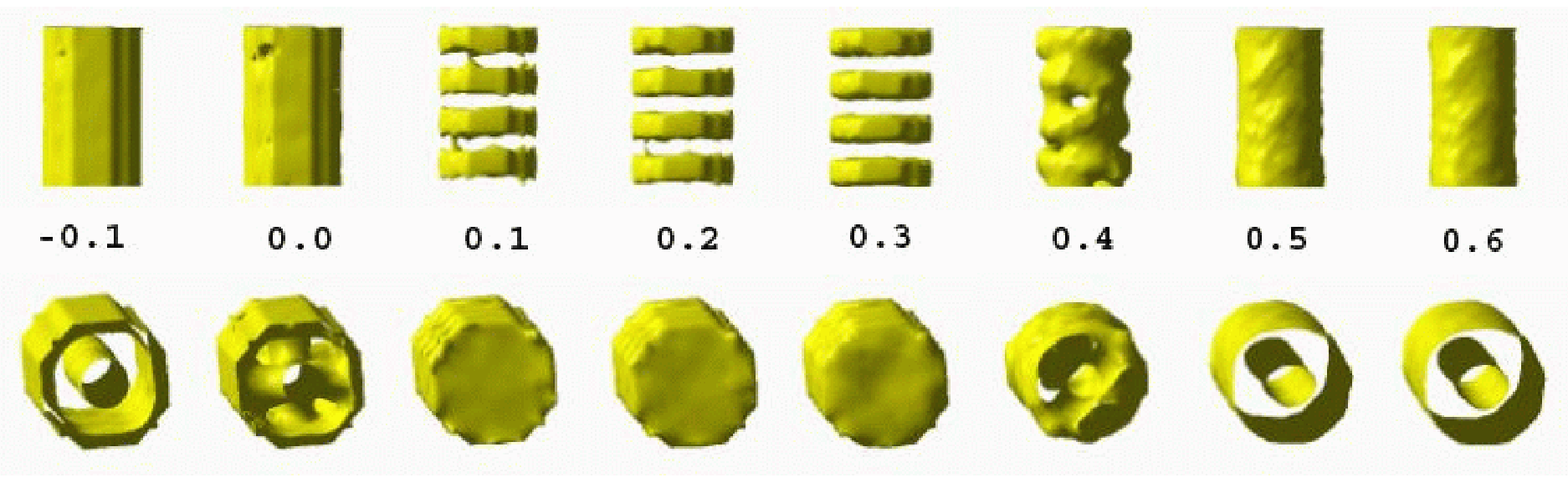,height=3.5cm}}
\put(1.9,4){\epsfig{file=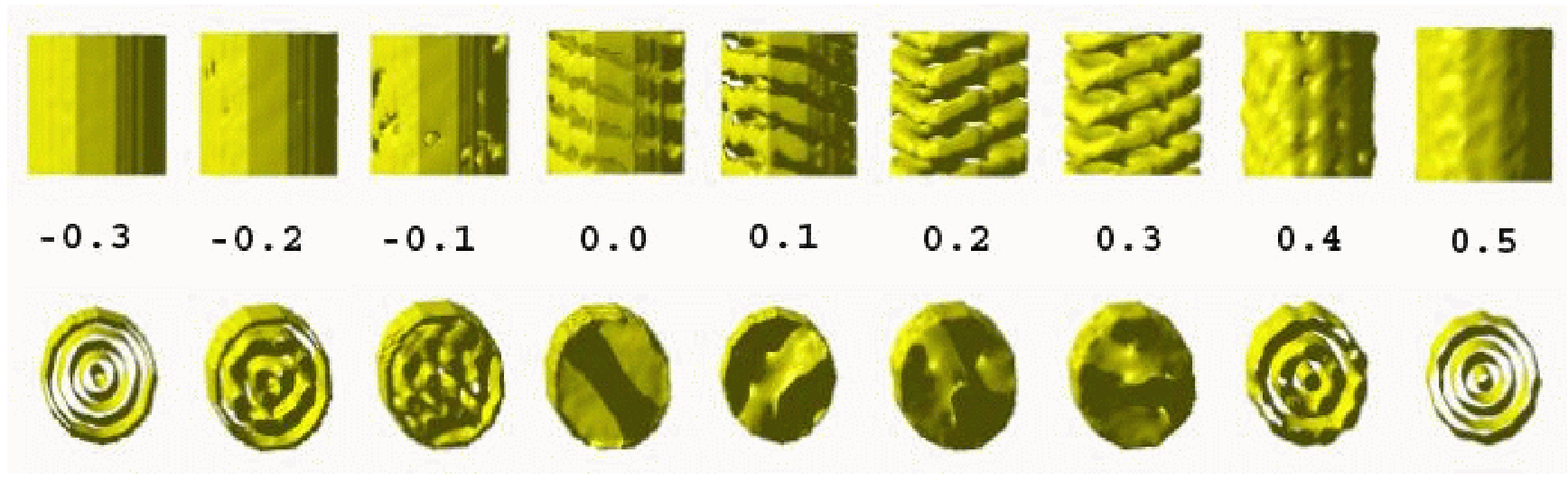,height=3.5cm}}
\put(1.8,0){\epsfig{file=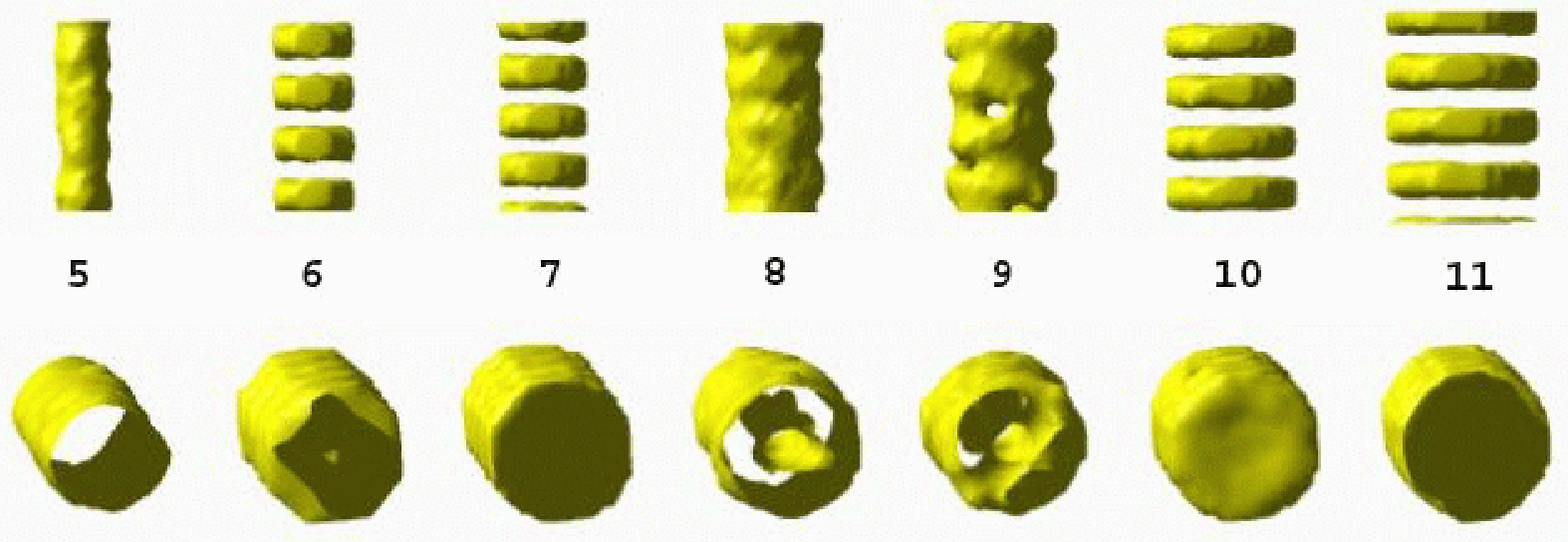,height=3.5cm}}
\put(1.5,9.2){a)}
\put(1.4,5.2){b)}
\put(1.3,1.2){c)}
\end{picture}
\caption{}
\label{fig13}
\end{figure}

\newpage

\begin{figure}
\setlength{\unitlength}{1cm}
\begin{picture}(10,10)
\put(0,0){\epsfig{file=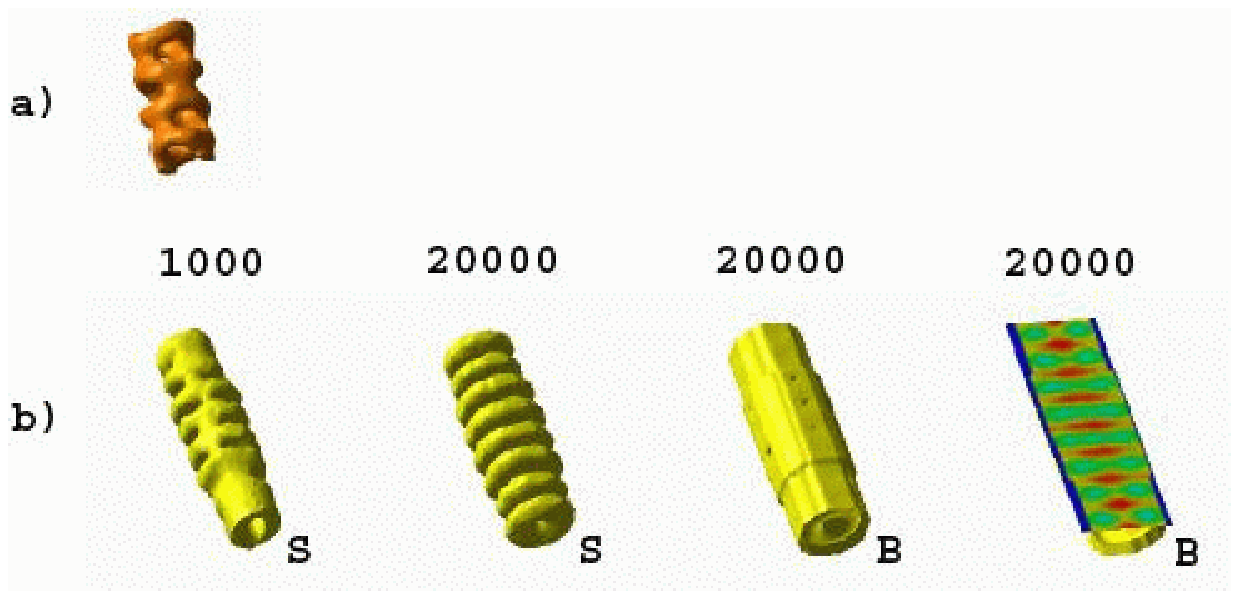,width=14cm}}
\end{picture}
\caption{}
\label{fig14}
\end{figure}

\end{onecolumn}


\begin{thebibliography}{10}

\bibitem{wangbook} Q. Wang, In: {\it Nanostructured Soft Matter: Experiment, Theory, Simulation 
and Perspectives}, Ed. A.V. Zvelindovsky, 498-528 (Springer, Dordrecht, 2007). 

\bibitem{fre99} F.~S.~Bates and G.~H.~Fredrickson, Physics Today, {\bf
52}, 32 (1999).

\bibitem{matsen98} M.W. Matsen, Curr. Opin. Colloid Interface Sci. {\bf 3}, 40 (1998).

\bibitem{binder99} K. Binder, Adv. Polym. Sci. {\bf 138}, 1 (1999).

\bibitem{fasolka01} M.J. Fasolka and A.M. Mayes, Annu. Rev. Mater. Res. {\bf 31}, 323 (2001).

\bibitem{hashimotobook} T. Hashimoto, In: {\it Nanostructured Soft Matter: Experiment, Theory, Simulation 
and Perspectives}, Ed. A.V. Zvelindovsky, 45-98 (Springer, Dordrecht, 2007). 

\bibitem{radzilowski96} L.H. Radzilowski and B.L. Carvalho and E.L. Thomas, J. Polym. Sci. Part B
Polym. Phys. {\bf 34}, 3081 (1996). 

\bibitem{huinink00} H.P. Huinink and J.C.M. Brokken-Zijp and M.A. van Dijk and G.J.A. Sevink, 
J. Chem. Phys. {\bf 112}, 2452 (2000).

\bibitem{wang01} Q. Wang and P.F. Nealy and J.J. de Pablo, Macromolecules {\bf 34}, 3458 (2001).

\bibitem{our_natmat} A. Knoll and K.S. Lyakhova and A. Horvat and G. Krausch and 
G.J.A. Sevink and A.V. Zvelindovsky and R. Magerle, Nat. Mat. {\bf 3}, 886 (2004).

\bibitem{ting} T. Xu and A.V. Zvelindovsky and G.J.A. Sevink and K.S. Lyakhova and H. Jinnai
and T.P.  Russell, Macromolecules {\bf 38}, 10788 (2005).

\bibitem{he01} X. He, M. Song, H. Liang, C. Pan, J. Chem. Phys., {\bf 114}, 10510-10513 (2001).

\bibitem{Sevpore} G.J.A. Sevink and A.V. Zvelindovsky and J.G.E.M. Fraaije and H.P. Huinink,
J. Chem. Phys., {\bf 115}, 8226 (2001). 

\bibitem{our_prl} A. Knoll and A. Horvat and K.S. Lyakhova and G. Krausch and 
G.J.A. Sevink and A.V. Zvelindovsky and R. Magerle, PRL {\bf 89}, 035501 (2002).

\bibitem{fasolka00} M.J. Fasolka and P. Banerjee and A.M. Mayes and G. Pickett and A.C. Balazs, Macromolecules {\bf 33}, 5702 (2000).

\bibitem{shin04} K. Shin, H.Q. Xiang, S.I.  Moon, T. Kim, 
T.J. McCarthy, T.P. Russell, Science, {\bf 306}, 76 (2004).

\bibitem{xiang04} H.Q. Xiang, K. Shin, T. Kim, S.I.  Moon,  
T.J. McCarthy and T.P. Russell, Macromolecules, {\bf 37}, 5660-5664 (2004).

\bibitem{wu04} Y.Y. Wu, G.S. Cheng, K. Katsov, S.W. Sides, J.F. Wang, 
J. Tang, G.H. Fredrickson, M. Moskovits, G.D. Stucky,
Nature Materials, {\bf 3}, 816 (2004).

\bibitem{xiang05} H.Q. Xiang, K. Shin, T. Kim, S.I.  Moon,  
T.J. McCarthy and T.P. Russell, J. Pol. Sci: part B: Pol. Phys., {\bf 43}, 3377 (2005). 

\bibitem{sun05} Y. Sun and M. Steinhart and D. Zschech and R. Adhikari and G.H. Michler and U. G\"osele, Macrom. Rap. Comm., {\bf 26}, 369-375 (2005). 

\bibitem{li06} W. Li and R.A. Wickham and R.A. Garbary, Macromolecules, {\bf 39}, 806 (2006).

\bibitem{chen06} P. Chen and X. He and H. Liang, J. Chem. Phys., {\bf 124}, 104906 (2006).

\bibitem{feng06a} J. Feng and E. Ruckenstein, Macromolecules, {\bf 39}, 4899 (2006).

\bibitem{feng06b} J. Feng and E. Ruckenstein, J. Chem. Phys., {\bf 125}, 164911 (2006).

\bibitem{wang07} Q. Wang, J. Chem. Phys. {\bf 126}, 024903 (2007).

\bibitem{THE} I.~W.~Hamley, {\it The physics of block copolymers}
(Oxford Univ. Press, Oxford, 1998).

\bibitem{fraaije97} J.~G.~E.~M.~Fraaije and B.~A.~C. van Vlimmeren and 
N.~M.~Maurits and M.~Postma and O.~A.~Evers and C.~Hoffmann and P.~Altevogt
and G.~Goldbeck-Wood, J. Chem. Phys., {bf 106}, 4260 (1996).

\bibitem{vlimmer99}  B.~A.~C.~van Vlimmeren, N.~M.~Maurits, A.~V.~Zvelindovsky,
G.~J.~A.~Sevink, and J.~G.~E.~M.~Fraaije, Macromolecules, {\bf 32}, 646
(1999).

\bibitem{matS}  M.~W.~Matsen and M.~Schick, Phys. Rev. Lett. {\bf 72}, 2660
(1994).

\bibitem{sevink99} G.~J.~A.~Sevink and A.~V.~Zvelindovsky and 
B.~A.~C. van Vlimmeren and N.~M.~Maurits and J.~G.~E.~M. Fraaije, 
J. Chem. Phys., {\bf 110}, 2250 (1999).

\bibitem{maurits96a} N.M. Maurits and J.G.E.M. Fraaije and P. Altevogt and O.A. Evers,
Comp. \& Theor. Pol. Sci. {\bf 6}, 1 (1996).

\bibitem{maurits97} N.M. Maurits and J.G.E.M. Fraaije,
J. Chem. Phys. {\bf 107}, 5879 (1997).

\bibitem{freire03} J.J. Freire, C. McBride. Macrom. The. and Sim., {\bf 12}, 237 (2003).

\bibitem{sevink04} G.~J.~A.~Sevink and A.~V.~Zvelindovsky,  J. Chem. Phys. {\bf 121}, 3864 (2004).

\bibitem{wang00} G. Wang, G. Yan, P.F. Naeley and J.J. de Pablo, J. Chem. Phys. {\bf 112}, 450 (2000).

\bibitem{morita04} H. Morita and T. Kawakatsu and M. Doi and D. Yamaguchi and 
M. Takenaka and T. Hashimoto, J. Phys. Soc. Japan, {\bf 73}, 1371 (2004).

\bibitem{huinink01} H.P. Huinink and J.C.M. Brokken-Zijp and M.A. van Dijk and G.J.A. Sevink, 
Macromolecules {\bf 34}, 5325 (2001).

\bibitem{lyakhova06}  K.S. Lyakhova and A.V. Zvelindovsky and G.J.A. Sevink,
Macromolecules {\bf 39}, 3024 (2006).

\bibitem{ma06} M.L. Ma and V. Krikorian and J.H. Yu and E.L. Thomas and G.C. Rutledge,
Nanoletters {\bf 6}, 2969 (2006).

\bibitem{kyrylyuk06} A.V. Kyrylyuk and J.G.E.M. Fraaije, J. Chem. Phys. {\bf 125}, 164716 (2006).

\bibitem{owens89} J.N. Owens and I.S. Gancarz and J.T. Koberstein and T.P. Russell, 
Macromolecules {\bf 22}, 3380 (1989).

\bibitem{fengdpd06} J. Feng and H. Liu and Y. Hu. Macromol. Theory Sim., {\bf 15}, 674 (2006).

\end{thebibliography}
\end{document}